\documentclass[12pt, a4paper]{article}
\usepackage[left=2.5cm, right=2.5cm, top=2.5cm, bottom=2.5cm]{geometry}
\usepackage{graphicx}
\usepackage{amssymb, amsmath}
\usepackage{mathtools}
\usepackage{bbm}
\usepackage{natbib}
\usepackage{authblk}

\newcommand{\one}{\mathbbm{1}}

\def\T{{ \mathsf{\scriptscriptstyle{T}} }}
\begin{document}

\title{Smooth hazards with multiple time scales\\[2ex]
  }
\author{Angela Carollo\authorcr\small
        \texttt{carollo@demogr.mpg.de}\authorcr
        Max Planck Institute for Demographic Research\authorcr
        Rostock, Germany \authorcr
				LUMC, Leiden, The Netherlands
        \and
        Paul H.C.~Eilers\authorcr\small
        \texttt{p.eilers@erasmusmc.nl}\authorcr
        Erasmus MC, Rotterdam, The Netherlands
        \and
        Hein Putter\authorcr\small
        \texttt{h.putter@lumc.nl}\authorcr
        LUMC, Leiden, The Netherlands
        \and
        Jutta Gampe\authorcr\small
        \texttt{gampe@demogr.mpg.de}\authorcr
         Max Planck Institute for Demographic Research\authorcr
        Rostock, Germany
}

\maketitle

\begin{abstract}
\noindent  
Hazard models are the most commonly used tool to analyse time-to-event data. If more than one time scale is relevant for the event under study, models are required that can incorporate the dependence of a hazard along two (or more) time scales. Such models should be flexible to capture the joint influence of several times scales and nonparametric smoothing techniques are obvious candidates. $P$-splines offer a flexible way to specify such hazard surfaces, and estimation is achieved by maximizing a penalized Poisson likelihood. Standard observations schemes, such as right-censoring and left-truncation, can be accommodated in a straightforward manner. The model can be extended to proportional hazards regression with a baseline hazard varying over two scales. Generalized linear array model (GLAM) algorithms allow efficient computations, which are implemented in a companion R-package. 

\vspace*{0.5cm}
\sc{Keywords}: \rm {Time scales; multidimensional hazard; $P$-splines; GLAM algorithms}
\end{abstract}
\vskip 3ex

\section{Introduction}
\label{sect:introduction}

In survival analysis we model the duration from a time origin until an event of interest occurs. In many applications, however, several time scales can be of interest.  In clinical examples time since disease onset or time since start of treatment are significant time scales, but also the patient's age, which is time since birth, can be a relevant scale.

Time itself is not the cause of events, but describes `the scale along which other causes operate' \citep{Berzuini:1994}, and time can be measured with respect to different origins thereby defining several time scales. These time scales serve as proxies for different underlying causes or factors that are difficult or impossible to measure otherwise.
For example, time since disease onset can be a proxy for the biological mechanism of the disease progression, while time since treatment can illustrate the cumulative effect of the therapy, and the age of the patient can represent the changing capacity to resist (co)morbidity load. The aspects captured by different time scales may well operate nonlinearly and interact with each other. Also, the effects of covariates may differ depending on the time scale used.

Most commonly time-to-event data are analyzed by means of hazard models. Several strategies for handling two (or more) time scales in survival analysis were proposed in the literature.

The simplest approach is to select a single time scale that is deemed most appropriate and along which the (baseline) hazard changes.  Covariates then modify this hazard. The specific mode of action depends on the particular model, such as proportional or additive hazards. Including (an)other time scale(s) in such models is through time-varying covariate(s).

\citet{Thiebaut:2004} recommend to use age rather than time-on-study as time scale for the analysis of epidemiological cohort studies. In contrast, \citet{Pencina:2007} found, in simulation studies, that models with time-on-study as the main time scale are often more reliable than age-scale models unadjusted for age at entry. A similar suggestion is made by \citet{Chalise:2013}.
When modelling time-varying environmental exposure, \citet{Griffin:2012} found that controlling for calendar time in models where the exposure is highly correlated with calendar time, significantly decreases the performance of such models, because of collinearity issues.
However, for a similar problem, \citet{Wolkewitz:2016} suggested that calendar time should be included as a covariate in models where time-since-admission to the hospital is the main time scale.
Obviously, there is no consensus about which time scale should be preferred when estimating hazard models in case multiple time scales are involved and choices are highly application-specific.

A different strategy, which is mostly found in reliability applications, is to compose an `ideal time scale' \citep{Duchesne:1999} as a combination of the others, mainly by dimension reduction techniques. An early example is \citet{Farewell:1979}, further discussion was presented in \citet{Oakes:1995}. Similar approaches of dimensionality reduction were proposed in \citet{Kordonsky:1997} and \citet{Duchesneetal:2000, Duchesne:2002}. 

The explicit consideration of rates over two time scales simultaneously dates back to \citet{Lexis} who introduced the device that we now call the Lexis diagram. \citet{Keiding:1990} provides an extensive review, including the history of the Lexis diagram. 
\citet{Efron:2002} proposed a two-way hazard model in which the log-hazard is expressed as the sum of two baseline hazards, one for each time scale, and a linear predictor term for the covariates. 
The two one-dimensional log-hazards are specified parametrically. 
\citet{Iacobelli:2013} use the log-additive two-way hazard model to estimate transitions rates in an illness-death model but, to relax rigid parametrizations, use spline functions for the two univariate log-hazards. The log-additive structure of the two-way hazard model implies that the shape of the hazard along one axis is multiplied by the value of the hazard at the other axis so that this basic shape is preserved. If the two time scales interact beyond such proportionality a more flexible model is needed.

\citet{Scheike:2001} proposed an additive hazard model for two time scales. Covariate effects are modeled by an additive Aalen model on each of the time scales, their sum forms the overall (univariate) hazard. A Bayesian non-parametric approach was proposed by \citet{Harkanen:2017}, in which the Lexis plane is divided into strips. A piecewise constant hazard model is fitted in each of these strips and   through the specification of the prior some smoothing is achieved within and across the strips.

We propose multidimensional $P$-spline smoothing \citep{EilersMarx:2021} to estimate a smooth hazard over two time scales. For that purpose the data are split in small two-dimensional bins (squares or rectangles) of equal size and events and times-at-risk within these bins are determined. This allows to exploit the well-known correspondence between hazard estimation and Poisson regression. The logarithm of the hazard surface is expressed as linear combination of tensor products of $B$-splines, and the spline coefficients are restrained by roughness penalties that can operate differently along the two axes for anisotropic smoothing. This specification allows to capture interactions in two dimensions. The approach can be extended to proportional hazards (PH) regression in which covariates modify a baseline risk surface. The binning of the data makes this approach a generalized linear array model (GLAM) for which efficient algorithms are available, see \citet{Currie:2006}. 

In Section~\ref{sec:2TS} we introduce the basic definitions for the hazard model over two time scales and the data example that we are going to analyze in this paper. In Section~\ref{sec:smoothhaz} we describe hazard estimation via $P$-splines, first in the case of a single time scale and then for a hazard that varies over two time scales. We contrast the results of the one- and two-dimensional hazard for the data example. In Section~\ref{sec:2DPH} we extend the model to proportional hazards regression and Section~\ref{sec:sim} presents a simulation study. The data are reanalyzed in a PH model in Section~\ref{sec:colonPH}, and we conclude with a discussion. Some computational details are presented in the Appendix.

\section{Hazard functions over multiple time scales}\label{sec:2TS}
\begin{figure}
  \includegraphics[width=0.95\textwidth]{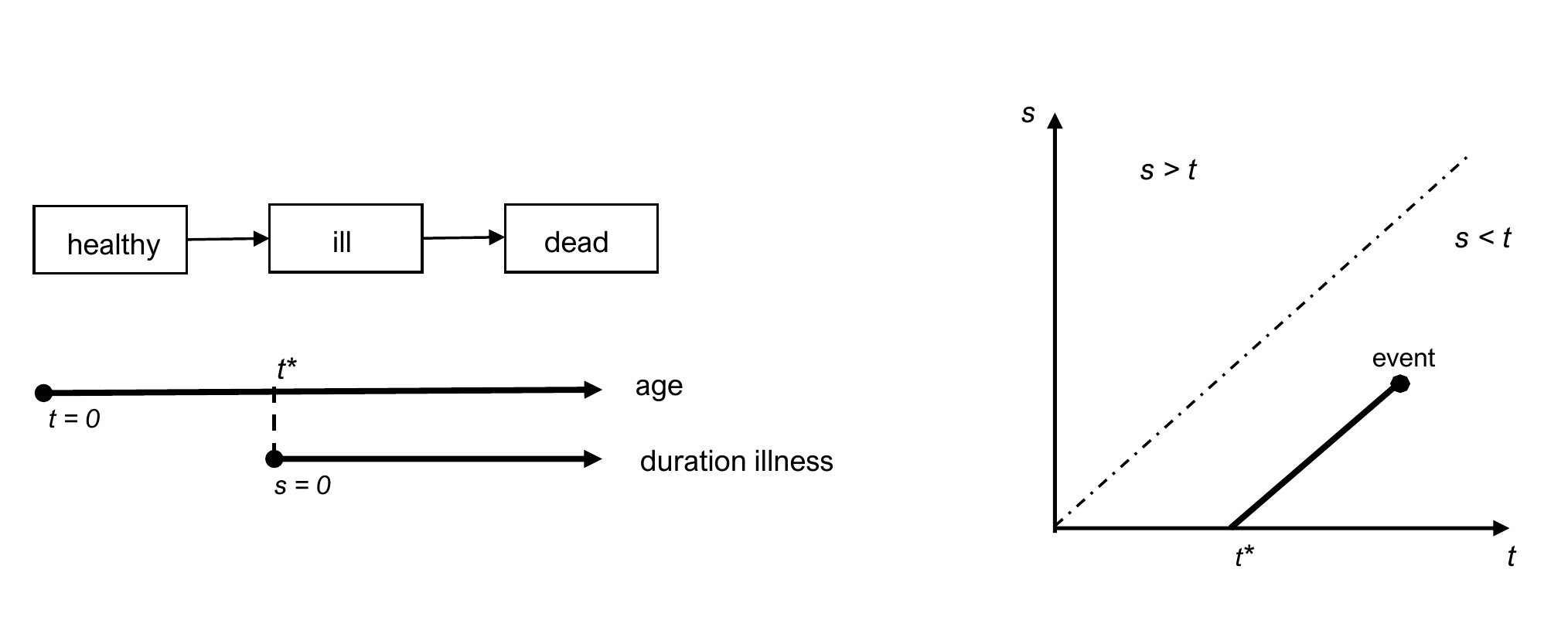}
  \caption{Simple illness-death model with two time scales.}\label{fig:2TS}
\end{figure}

Multiple time scales differ in their origin but time progresses at  the same speed along all time scales. As a specific example, consider a simple illness-death model in which patients move from the `healthy' state to the state `ill' and the event of interest, for which the hazard is to be modelled, is death, see Figure~\ref{fig:2TS}. If the first time scale $t$ denotes the age of the patient (with origin either birth or, for late-onset diseases, a later appropriate age before which the disease does not occur) then a second time scale $s$ that identifies the duration of illness starts at entry into the state `ill'. The difference in the origins of the time scales $t$ and $s$ is given by the age $t^*$ at which the patient falls ill. This difference will vary between individuals.

We can portray individual trajectories in the Lexis diagram (Figure~\ref{fig:2TS}, right).
If time is measured in the same unit for both axes, so that an increment of
$\varepsilon$ in $t$ corresponds to the same increment in $s$, then individuals move along diagonal lines with
slope 1. The individual lines start at $(t=t^*; s=0)$ and extend until the event (or loss to follow-up) occurs at, say, $(t=t^*+v; s=v)$. All trajectories are situated in the lower right open triangle for which $t>s$.

The hazard $\lambda(t, s)$ over the two time scales $t$ and $s$ gives the instantaneous risk of experiencing the event (death, in the example) at age $t$ and duration of the illness $s$, given that the individual is still alive at $(t,s)$. It is defined over the triangular domain $\{(t,s) \subset \mathbb R ^2 \,:\, t >s \}$.
If we denote the line between $(t,s)$ and $(t+\varepsilon, s+\varepsilon)$, along which individuals advance, by $L_\varepsilon (t,s ) =\{ (t+\varepsilon ; s+\varepsilon) : \varepsilon \geq 0\}$, then the hazard is formally defined as
\begin{equation}
\label{eq:haz2D}
\lambda(t, s) = \lim _{\varepsilon \downarrow 0}\; 
\frac{P\left \{ \;\text{event} \in  L_\varepsilon (t,s)\;|\; \text{no event before } (t, s) \; \right \} }{\varepsilon} .
\end{equation}
We will assume that $\lambda(t,s)$ is a smooth function over its domain. Estimation of  $\lambda(t,s)$ will be discussed in Section~\ref{sec:haz2D}.

For a given value $t^*$, when an individual enters the intermediate state and the second time scale commences, the individual progresses along the diagonal cutline of the hazard surface.
This perspective indicates that the values of $\lambda(t,s)$ can be obtained just as well if we consider the hazard along the second time axis $s$ but indexed over the value $u=t-s$, which equals the difference in the origins of the two scales $t$ and $s$. So this difference equals $t^*$, the value of $t$ for which the second scale takes off. The model can thus be equivalently interpreted as a one time scale model over $s$, where the hazard is smoothly modulated across the values of the variable $u$. So the alternative interpretation of 
\begin{equation}\label{eq:hazard2Dtrafo}
\breve\lambda(u, s) =\lambda(u+s, s)
\end{equation}
is a smooth interaction model between the duration-specific hazard and the, in this example, entry-age $u$. The hazard trajectory along $s$ can be different for different values of $u$, but the change is assumed to be gradual along the $u$-axis.

The transformation from $(t,s)$ to $(u,s)$ is linear and invertible, in matrix notation we obtain
\begin{equation}
  \label{eq:xtoy}
  \begin{pmatrix}
     u \\
    s 
  \end{pmatrix}
  =
  \begin{pmatrix}
    t - s\\
     s
   \end{pmatrix}
   =
 \begin{pmatrix}
    1&-1\\
    0&~~1
  \end{pmatrix} 
  \begin{pmatrix}
     t \\
    s 
  \end{pmatrix} .
\end{equation}
This change in perspective also changes the domain over which the hazard surface is defined: $(u,s) \in \mathbb R_+^2$, so the domain of $\breve \lambda(u,s)$ is the full positive plane. Obviously, there is a one-to-one correspondence between the two surfaces and smoothness of one implies smoothness of the other.

\subsection{Data example: Adjuvant therapies for colon cancer}\label{sec:colondata}
To illustrate the above we consider a dataset that we will revisit in Sections~\ref{sec:smoothhaz} and \ref{sec:colonPH}.
We will analyze data from a clinical trial on colon cancer and the effects  of two adjuvant therapies after colon resection \citep{Laurie:1989, Moertel:1995}. The data are  included in the R-package \texttt{survival} \citep{survival-package}. Patients were randomized (after recovery from surgery) into one of the two treatment groups or the control group (no treatment). The two treatments were either Levamisole, a drug showing immunostimulatory activity, or a combination of Levamisole and Fluorouracil, a moderately toxic chemotherapy agent. Patients were followed until death or censoring and can experience recurrence of the cancer during follow-up.
\cite{Moertel:1995} report that survival of patients treated with Levamisole alone is the same as those in the control group, while individuals treated with Levamisole + Fluorouracil experienced a better survival right after randomization. The combination therapy was also found to effectively reduce the recurrence rate. However, the authors also report that, after recurrence of the cancer, patients in this third group experience shorter survival times than patients in the control group and that survival after recurrence was clearly related to time at recurrence \citep{Moertel:1995}.

This observation motivates our analysis which studies mortality after recurrence and considers the two time scales $t$:\,`time since randomization' and $s$:\,`time since recurrence'.  The dataset contains 929 individuals, 468 (50.4\%) experienced a recurrence of the cancer during follow-up. Of those patients with recurrence seven left the risk set at the recorded time of recurrence, leaving 461 who were followed up further.  This subsample of 461 individuals, of whom 409 died during follow-up and 52 were right-censored, is the focus of our analysis. Additional covariates on the patients (sex, age at surgery) and on the charateristics of the tumor are available as well.

\begin{figure}%
  \includegraphics[width=\columnwidth]{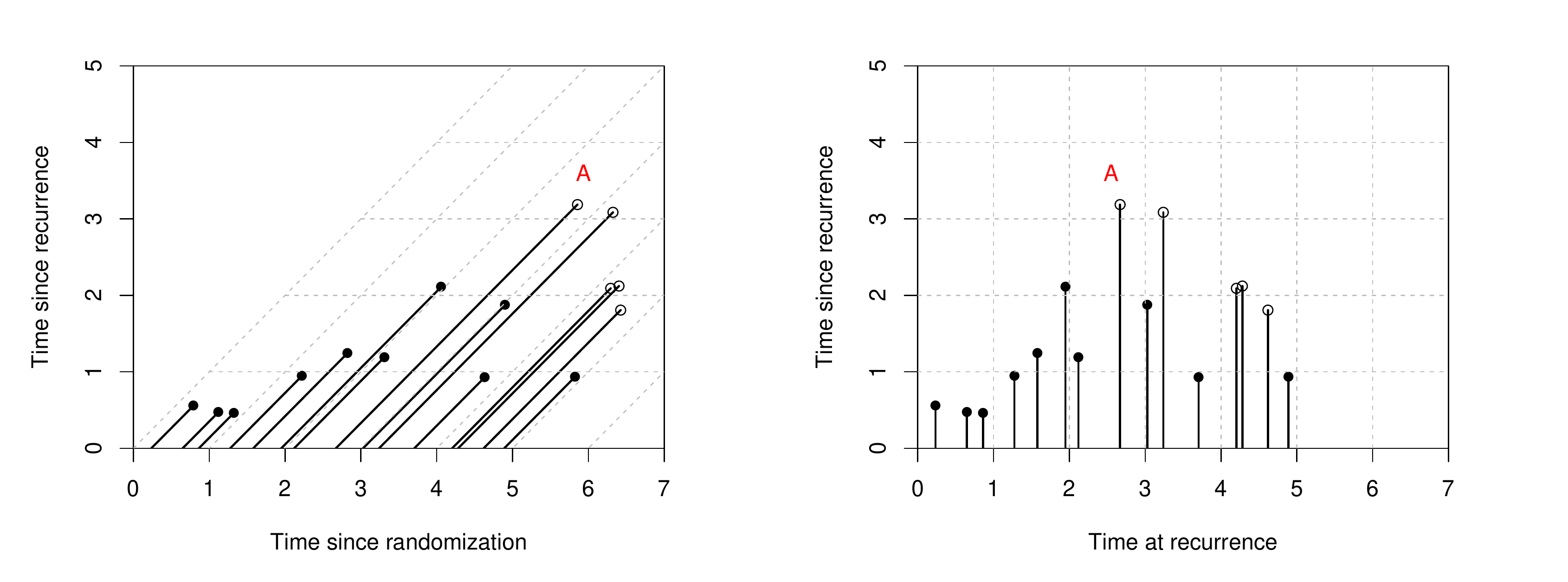}%
  \caption{\label{fig:transformation}~Colon cancer data, event death ($\bullet$) or censoring($\circ$): Trajectories of 15 randomly selected individuals in the $(t,s)$ Lexis-plane (left panel) and transformed trajectories in $(u=t-s, s)$ plane (right panel). One individual (A) is labeled in both displays.}
\end{figure}

Figure~\ref{fig:transformation} shows the trajectories of 15 randomly selected individuals from this dataset, in the left panel over the two time scales $t$ and $s$. In the right panel the same individuals are portrayed but this time according to the value $u$:\,`time at recurrence'
and $s$:\,`time since recurrence', and $u=t-s$, see equation~(\ref{eq:xtoy}).

\section{Smoothing hazards with $P$-splines}\label{sec:smoothhaz}
Before we describe the estimation of the two-dimensional hazard $\lambda(t,s)$, we briefly outline one-dimensional hazard smoothing with $P$-splines. This allows us to fix notation and to introduce the general principle, which is then extended in Section~\ref{sec:haz2D}. For now we ignore the potential effects of covariates, regression models will be addressed in Section~\ref{sec:2DPH}.

\subsection{Univariate hazard smoothing}\label{sec:estim1D}

Flexible hazard modelling can be achieved by splitting the time axis into $n_t$ bins $(\tau_{j-1}, \tau_j]$, $j=1, \ldots n_t$, and allowing a different hazard level $\lambda _j$ across bins. The resulting likelihood is equivalent to a Poisson model for the event counts $y_j$ in each bin \citep{Holford:1980, LairdOlivier:1981}, in which the expected values $\mu_j$ are the product of the hazard level $\lambda _j$ and the total time at risk $r_j$,
\begin{equation}\label{eq:Poisson}
y_j \sim \text{Poisson} (\mu _j)  \text{\quad with\quad} \mu_j = r_j \, \lambda _j, \quad j=1, \ldots, n_t.
\end{equation}
Each individual $i$, $i=1,\ldots , n$, in the sample contributes its at-risk time $r_{ij}$ in the bins and the bin-specific event-indicators $y_{ij}$, which equal 1, if an event occurred in bin $j$, and zero otherwise. Hence $y_j= \sum_{i} y_{ij}$ and $r_j= \sum_{i} r_{ij}$. For right-censored observations $y_{ij}=0$ for all $j$, and left-truncated observations contribute positive at-risk times $r_{ij}$ only after their time of entry into the study. 
The canonical parameter in model (\ref{eq:Poisson}) is $\ln \mu_j = \ln r_j + \ln \lambda_j = \ln r_j + \eta_j$, the sum  of the log-hazard~$\eta_j$ and the known offset $\ln r_j$.  Maximum likelihood estimation here results in the common occurrence-exposure rates $\hat \lambda _j = e^{\hat \eta_j} = y_j/r_j$.

Choosing a large number of bins may allow more flexible hazards but inevitably the resulting estimates will show erratic behavior in areas where only few individuals are observed. In any case, a step function for $\eta (t)$, and consequently for $\lambda (t)$, is only a rough approximation to the (log-)hazard function that commonly is assumed to be smooth. These drawbacks can be overcome by $P$-spline smoothing \citep{Eilers:1996, vanHouwelingen:2000}. 

In the following we choose bins of equal length $h$ so that $\tau_j = j\cdot h$. The bin width $h$ will be relatively small and consequently the number of bins $n_t$ large. The bins are defined such that they cover the range of observed event or censoring times, respectively.

The log-hazard $\eta (t)$ is modeled as a linear combination of $B$-splines (of degree $p$) that are defined on a regular grid of knots so that
\begin{equation}\label{eq:lp}
\eta =(\eta_1, \ldots , \eta_{n_t})^\T =(\ln \lambda_1, \ldots , \ln \lambda_{n_t}) ^\T = B\alpha .
\end{equation}
The matrix $B = (b_{jl})$ is of dimension  $n_t \times c_t$ and contains the $c_t$  $B$-splines evaluated at the midpoints of the bins $m_j = \tau_j - h/2$, that is, $b_{jl} = B_l(m_j)$ where $B_l$ is the $l^{\text{th}}$ $B$-spline in the basis.
The number of $B$-splines $c_t$ can be relatively large because a roughness penalty on the coefficients $\alpha =(\alpha_1, \ldots , \alpha _{c_t}) ^\T $ will prevent overfitting.

The penalty is based on differences of order $d$ between neighbouring coefficients in $\alpha$. These differences can be calculated by multiplication with a matrix $D_d$ of dimension $(c_t -d)\times c_t$, which for $d=1$ and $d=2$ is as follows
\begin{equation*}
D_1 = 
\begin{pmatrix*}[r]
-1 &1& 0&\ldots&0\\
0&-1&1& \ldots&0\\
\vdots&&\ddots&\ddots&\vdots\\
0&\ldots&0& -1&1\\
\end{pmatrix*}
\hspace {4em}
D_2 = 
\begin{pmatrix*}[r]
1&-2&1&0&\ldots&0\\
0&1&-2&1&\ldots&0\\
\vdots&&\ddots&&\ddots&\\
0&\ldots&&1&-2&1\\
\end{pmatrix*} .
\end{equation*}
The sum of squares of the differences $\| D_d \,\alpha \| ^2 = \alpha^\T D_d^\T D_{d\,} \alpha$ provides a roughness measure and serves as the penalty. Larger values of $\| D_d \,\alpha \| ^2$ correspond to less smooth estimates.

The penalized log-likelihood of the unknown parameters $\alpha$, given the vectors of bin-wise event counts $y =(y_1,\ldots, y_{n_t})^\T$ and exposure times $r=(r_1,\ldots, r_{n_t})^\T$ is 
\begin{equation}\label{eq:logLpen}
\ell_\varrho(\alpha; y, r) = \sum_{j=1}^{n_t} \left (y_j \, \ln \mu_j - \mu_j \right ) \, - \,
\frac{\varrho}{2}\; \| D_d \,\alpha \| ^2 .
\end{equation}
The smoothing parameter $\varrho$ in (\ref{eq:logLpen}) balances the model fit, as expressed by the log-likelihood, and the smoothness of the estimates induced by the penalty. For a given value of $\varrho$, differentiation
of (\ref{eq:logLpen}) leads to the following system of equations 
\begin{equation}\label{eq:normaleq}
B^\T(y-\mu) = \varrho \, D^\T D \alpha ,
\end{equation}
which is solved by a (penalized) iteratively-weighted least-squares (IWLS) scheme
\begin{equation}\label{eq:IWLS}
(B^\T\tilde W B + \varrho \, D^\T D ) \,  \alpha = 
B^\T \tilde W B \tilde \alpha + B^\T \left ( y-\tilde \mu \right ).
\end{equation}
The tilde indicates the current value in the iteration, i.e., $\tilde \eta= B\tilde \alpha$ and $\tilde \mu = r \odot e^{\tilde \eta}$ (where~$\odot$ denotes elementwise multiplication). The weight matrix for the Poisson model is $\tilde W = \text{diag}(\tilde\mu)$. 

The optimal value of $\varrho$ can be obtained by minimizing AIC (Akaike's Information Criterion) over a grid, linear on log-scale, of $\varrho$-values
\begin{equation*}
\text{AIC}(\varrho) = \text{Dev}(\mu; y) + 2\, \text{ED} .
\end{equation*}
$\text{Dev}(\mu; y) = 2 \sum_{j} \, y_j \ln \frac{y_j}{\mu_j} $ is the Poisson deviance and the effective dimension $\text{ED}$ is obtained as trace of the hat matrix $H$
\begin{equation}\label{eq:hatmatrix}
\text{ED} = \text{tr}(H) = 
\text{tr}\left (B \,(B^\T\hat W B +\varrho \, D^\T D)^{-1} B^\T\hat W \right ) 
 = \text{tr}\left ( (B^\T\hat W B +\varrho \, D^\T D)^{-1} B^\T\hat W B \right ).
\end{equation}

The variance-covariance matrix of the coeffcients is 
$\text{Cov}(\hat \alpha) \approx (B^\T\hat W B +\varrho \, D^\T D)^{-1}$, from which 
$\text{Cov}(\hat \eta) =\text{Cov}(B \hat \alpha) = B\, \text{Cov}(\hat \alpha)\, B^\T $ results \citep[see][Appendix~F]{EilersMarx:2021}. 

Once the coefficients $\hat \alpha$ have been estimated the (log-)hazard can be obtained for additional values $\mathring t$, other than the midpoints $m_j$, by evaluating the $c_t$ $B$-splines at $\mathring t$ and obtaining
\begin{equation}
\hat \eta (\mathring t ) = (B_1(\mathring t), \ldots , B_{c_t}(\mathring t) ) \,\hat\alpha \text{\quad and\quad}
\hat \lambda ( \mathring t ) = \exp \{\hat \eta (\mathring t ) \} .
\end{equation}

\begin{figure}
\includegraphics[width=\textwidth]{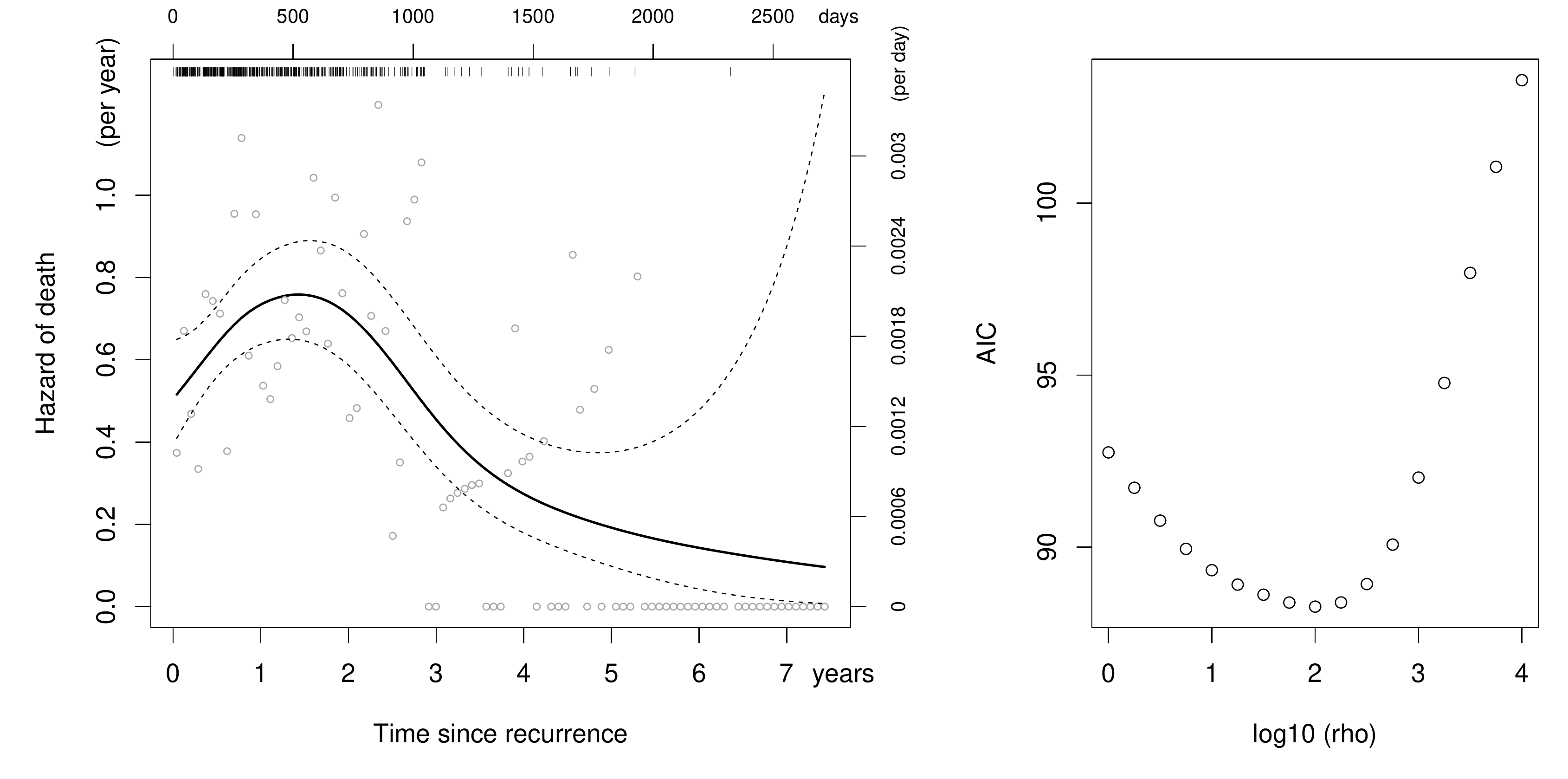}
\caption{Estimated hazard of death $\lambda(t)$ over $t$:\,\textsl{time since recurrence} for colon cancer data. Left: $P$-spline estimate for $n_t=91$ bins of length 30 days, $c_t=20$ cubic $B$-splines and penalty order $d=2$ (solid line). Observed event times are indicated at the top of the panel. Effective dimension ED=4.3. Grey circles give the bin-wise estimates $\hat \lambda _j = y_j/r_j$. Hazards are given per year.
Dashed lines represent $\exp\{ \hat \eta \pm 2\cdot s.e.(\hat \eta)\}$. 
Right: $\text{AIC}(\varrho)$ over a grid of values of 
$\log_{10}\varrho$.}\label{fig:1Dhazard}
\end{figure}

As an illustration we estimate the hazard of death after recurrence for the colon cancer data introduced in Section~\ref{sec:colondata}. The maximum follow-up time was $2,725$ days, which is about 7.5 years. We split the time axis in $n_t=91$ bins of length $30$ days (covering $2,730$ days). $c_t =20$ cubic $B$-splines were used (so $p=3$) and a second-order penalty ($d=2$). Figure \ref{fig:1Dhazard} shows the resulting estimate $\hat \lambda (t)$ and the AIC-profile, from which the optimal $\varrho$ was obtained as $\varrho_{opt}=10^2$.

\subsection{Smoothing two-dimensional hazard surfaces}
\label{sec:haz2D}

This approach to obtain smooth hazards can be extended to bivariate surfaces, in our case to estimate the smooth hazard $\breve \lambda (u,s)$, see (\ref{eq:hazard2Dtrafo}). The binning of the data now extends over the $u$-axis and the $s$-axis, leading to a tesselation of $n_u\times n_s$ squares (or rectangles, if different bin widths are chosen for the two axes). Again the bins can be narrow and hence $n_u$ and $n_s$ relatively large.
Each individual contributes to a vertical sequence of squares (see Figure~\ref{fig:transformation}, right), depending on the value of $u$. 

Instead of the $n_t$-vectors of events $y_{ij}$ and exposures $r_{ij}$ we now have, for each individual~$i$, $n_u\times n_s$ matrices $Y^i_{jk}$ and $R^i_{jk}$, both sparsely filled, that contain the individual event and exposure information. When no additional covariates are included we again can sum over all individuals and obtain the $n_u\times n_s$ matrices of event counts $Y=(y_{jk})$ and of times at risk $R=(r_{jk})$. Following the same reasoning as in~(\ref{eq:Poisson}) we have 
\begin{equation}\label{eq:Poisson2}
Y \sim \text{Poisson}(M) \text{\qquad with\qquad} M=(\mu_{jk}) =R\odot \Lambda ,
\end{equation}
where $ \Lambda =(\breve \lambda _{jk})$ contains the hazard levels over the two-dimensional bins.

$P$-spline smoothing in two dimensions can be achieved by using tensor products of $B$-splines and a two-dimensional penalty. This was introduced in \citet{EilersMarx:2003}. \citet{Currie:2004} employ the approach to smooth and forecast mortality tables, a related R-package is described in \citet{Camarda:2012}.

For each time axis a separate marginal $B$-spline matrix is constructed (see Section~\ref{sec:estim1D}) that we denote by $B_{u} \in \mathbb R ^{n_u\times c_u}$ and $B_{s} \in \mathbb R ^{n_s\times c_s}$, respectively. $c_u$ and $c_s$ are the numbers of $B$-splines used for each axis. The rows of $B_u$ and $B_s$ contain the $B$-splines  evaluated at the mid-points of the respective marginal bins.
The regression matrix $B$ for the two-dimensional log-hazard $\breve\eta(u,s) = \ln \breve \lambda (u,s)$ is then defined as the tensor product 
\begin{equation}\label{eq:tensor}
B = B_s \otimes B_u
\end{equation}
and is of dimension $n_u n_s \times c_uc_s$. Here $\otimes$ denotes the Kronecker product.

The $c_u c_s$ regression coefficients $\alpha _{lm}$ are best arranged in the coefficient matrix $A=(\alpha_{lm})$.
If we arrange the $\breve\eta _{jk}$, which represent the log-hazard evaluated at the midpoints of the two-dimensional bins, correspondingly as $E =(\breve\eta_{jk}) = (\ln \breve\lambda_{jk}) \in \mathbb R^{n_u\times n_s}$, then we can express the linear predictor for the Poisson regression model (\ref{eq:Poisson2}) in vectorized form. We define $\eta = \text{vec}(E)$, which is of length $n_u n_s$, 
$\alpha = \text{vec}(A)$ of length $c_uc_s$ and therewith
\begin{equation}
\eta = B \, \alpha ,
\label{eq:2Dlp}
\end{equation}
which underpins the correspondence to (\ref{eq:lp}). Again a roughness penalty will be introduced so that also two-dimensional hazard smoothing is solved by penalized Poisson regression.

The penalty on the regression coefficients in $A$ also extends over two dimensions, one over the rows of $A$ and the other over the columns of $A$. The amount of smoothing in the two directions (along the two time axes) can be different to allow anisotropic smoothing. 
If $I_u$ and $I_s$ denote identity matrices of dimension $c_s$ and $c_u$, respectively, and $D_u$ and $D_s$ the difference matrices for the coefficients along $u$ and $s$, then the overall penalty matrix~$P$ of dimension $c_u c_s \times c_u c_s$ is obtained as the sum of two terms: One for the coeffcients in the direction of the rows and one in the direction of the columns of $A$
\begin{equation}
\label{eq:2Dpenalty}
	P = \varrho_u (I_s \otimes D_u^\T D_u) + \varrho_s (D_s^\T D_s \otimes I_u),
\end{equation}
where $\varrho_u$ and $\varrho_s$ are the smoothing parameters. The order of the differences (which was dropped in the notation) in the two parts of $P$ in principle can be different, although in many applications the same value is chosen.
The matrix $P$ takes the role of the single $\varrho D^\T D$ in equation~(\ref{eq:normaleq}).

The vectorized form in (\ref{eq:2Dlp}) stresses the correspondence to the one-dimensional set-up, however,
solving the IWLS equations in this form is computationally inefficient. As the data are on a regular grid this is a so called  generalized linear array model (GLAM; \citealp{Currie:2006}) for which skillful rearrangements allow a considerable gain in computation speed and memory use \citep{Eilers:2006}.
We rather write (\ref{eq:2Dlp}) as 
\begin{equation}\label{eq:2Dlp2}
E= B_u\, A\, B_s^\T ,
\end{equation}
replacing the large tensor product $B$ by products of smaller matrices and apply the GLAM algorithm. The procedure is outlined in Appendix~\ref{app:GLAM}.

Again, optimal values for the two smoothing parameters can be obtained by varying  $\log_{10}\varrho_u$ and $\log_{10}\varrho_s$ over a grid and choosing the combination leading to the smallest value of AIC. As fitting the model for all values on the $\varrho$-grid can be cumbersome, even when using the GLAM algorithm, numerical minimization of AIC is a good alternative. We follow \citet{EilersMarx:2021} and use \texttt{ucminf} from the R-package with the same name  \citep{ucminf}.

Once the coefficients $ \hat A =(\hat{\alpha}_{lm}) $ are computed, we can obtain estimates for the two-dimensional log-hazard  $\eta=\ln \lambda$ at arbitrary points $(\mathring t, \mathring s) = (\mathring u + \mathring s, \mathring s)$ by evaluating the marginal bases in (\ref{eq:2Dlp2}) at $\mathring u = \mathring t - \mathring s$ and $\mathring s$, respectively, and inserting 
\begin{equation*}
\hat{\eta}(\mathring t,\mathring s) = \hat{\breve\eta}(\mathring u ,\mathring s ) =  B_u (\mathring u ) \,\hat{A} \, B_s (\mathring s )^\T .
\end{equation*} 

The $c_u c_s \times c_u c_s$ variance-covariance matrix of the coefficients $\hat \alpha _{lm}$ is obtained, as in the one-dimensional case, as 
\begin{equation*}
\text{Cov}(\alpha) \approx (B^\T \hat W B + P )^{-1} 
\end{equation*}
with $B$ defined in (\ref{eq:tensor}) and $P$ in (\ref{eq:2Dpenalty}). The GLAM structure also facilitates the calculation of the variances of the log-hazard values in $\hat E= ( \hat{\breve\eta} _{jk})$. The details are outlined in Appendix~\ref{app:GLAM} as well.  

\begin{figure}[htbp]
\centering
\includegraphics[width=0.9\textwidth]{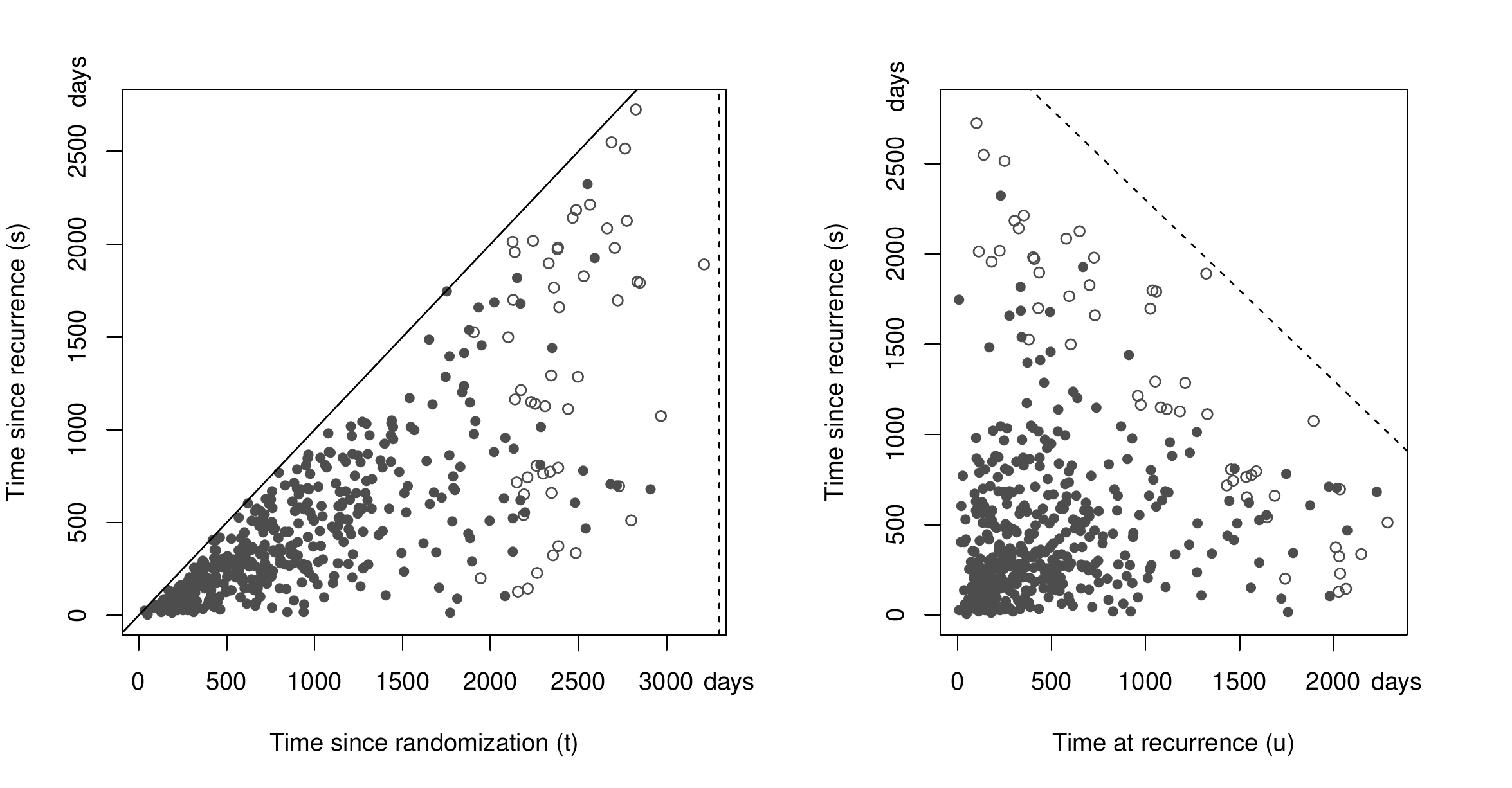}
\caption{Left: Distribution of exit times in $t$=`days since randomization' and $s$=`days since recurrence' for $n=461$ patients ($\bullet$ for deaths, $\circ$ for right-censored observations). The dashed vertical line marks the maximum follow-up time $t_{\max}= 3,214$ days since randomization. Right: Days after recurrence $s$ over time at recurrence $u =t-s$. Again the dashed line, $s=t_{\max} -u$, marks the area beyond which no data are observed due to end of follow-up.} \label{fig:colondescript}
\end{figure}

\subsection{Hazard of death along two time scales for colon cancer patients}\label{sec:colon2D}

For patients in the colon cancer study who experienced a recurrence the two time scales are $t$: `time since randomization' and $s$: `time since recurrence'. The sample size is $n=461$, of whom 409 died and 52 were alive at end of follow-up. Figure~\ref{fig:colondescript} shows the bivariate distribution of times at death or censoring in the $(t,s)$-plane and also as $u$: `time at recurrence' and $s$, from which we will estimate the two-dimensional hazard $\breve \lambda (u,s)$.

The maximum follow-up time $t_{\max}$ is $3,214$ days since randomization (about 8.8 years). Should hazard estimates be presented for times considerably beyond $t_{\max}$, then we clearly extrapolate. Extrapolation with 
$P$-splines is possible due to the penalty on the coefficients. In areas where no individuals are at risk the observations (bins) have zero weights and the penalty smoothly extends the coefficients in such areas \citep[see][]{Currie:2004}. Nevertheless extrapolation in areas not supported by any data should be applied cautiously. The extrapolation area $t>t_{\max}$, marked by a dashed line in Figure~\ref{fig:colondescript}, left, corresponds to the area above the dashed line $s=t_{\max} - u $ in  Figure~\ref{fig:colondescript}, right.

\begin{figure}[htbp]
\centering
\includegraphics[width=0.8\textwidth]{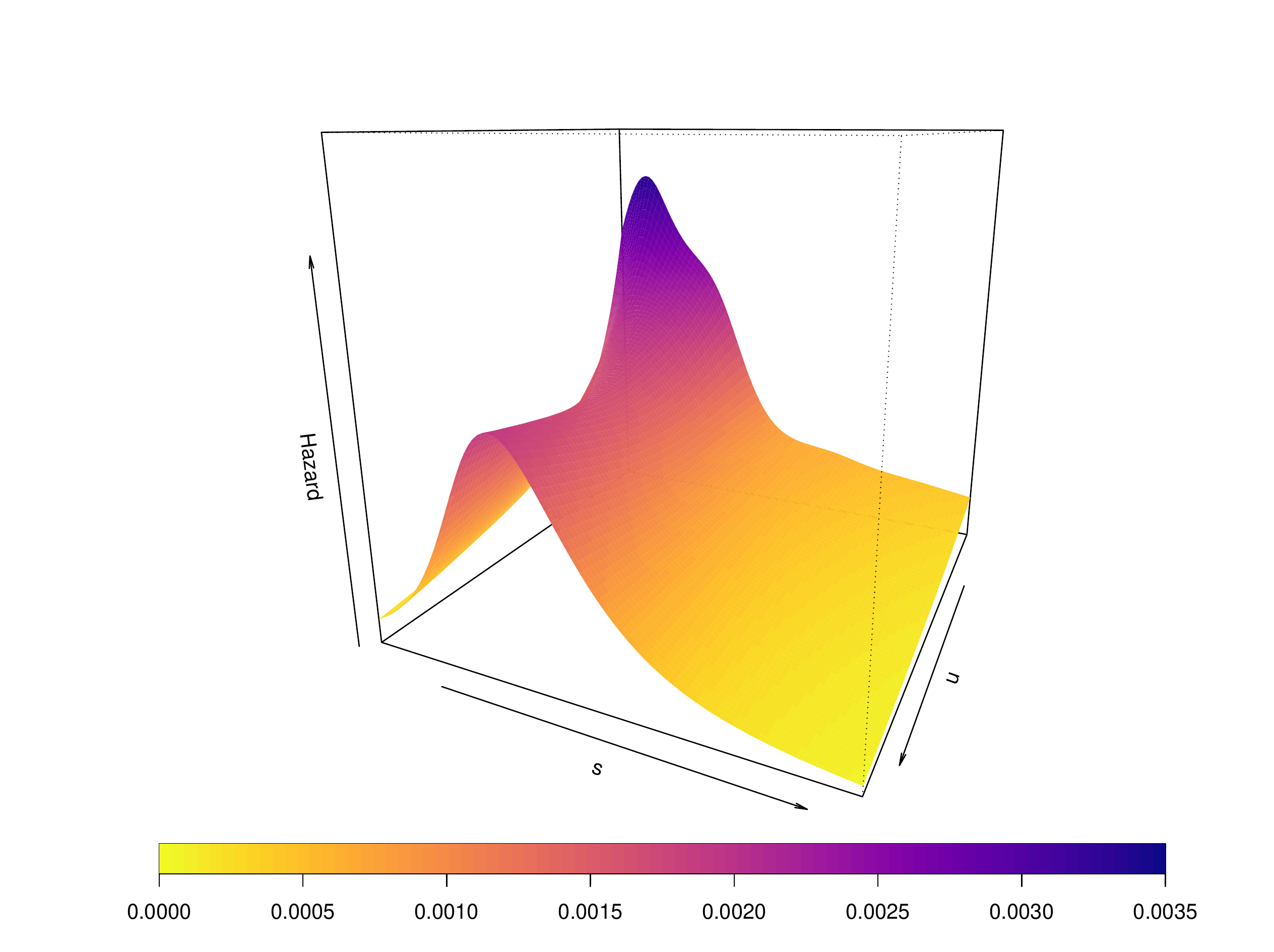}
\caption{Estimated hazard surface $\hat{\breve \lambda} (u,s)$. Cubic $B$-splines with 20 segments per axis, penalty order $d=2$. Smoothing parameters $\rho_u =10 ^{2.4}, \rho_s =10^{0.3}$ minimize AIC. Effective dimension ED=11.2}\label{fig:colon3Destim}
\end{figure}

To estimate the hazard surface $\breve \lambda (u,s)$ we cut the $(u,s)$-plane in bins (squares) of size $30$ by $30$ days. This implies $n_u=77 $ and $n_s = 91$. For each marginal basis cubic $B$-splines with 20 segments were used, so that $c_u=c_s=23$ and a total of $23^2= 529$ parameters $\alpha _{lm}$ have to be estimated. The order of the penalty was $d=2$ along both dimensions.
The optimal smoothing parameters were chosen by minimizing the $\text{AIC}(\rho_u, \rho_s)$.

\begin{figure}[tbp]
\centering
\includegraphics[width=0.44\textwidth]{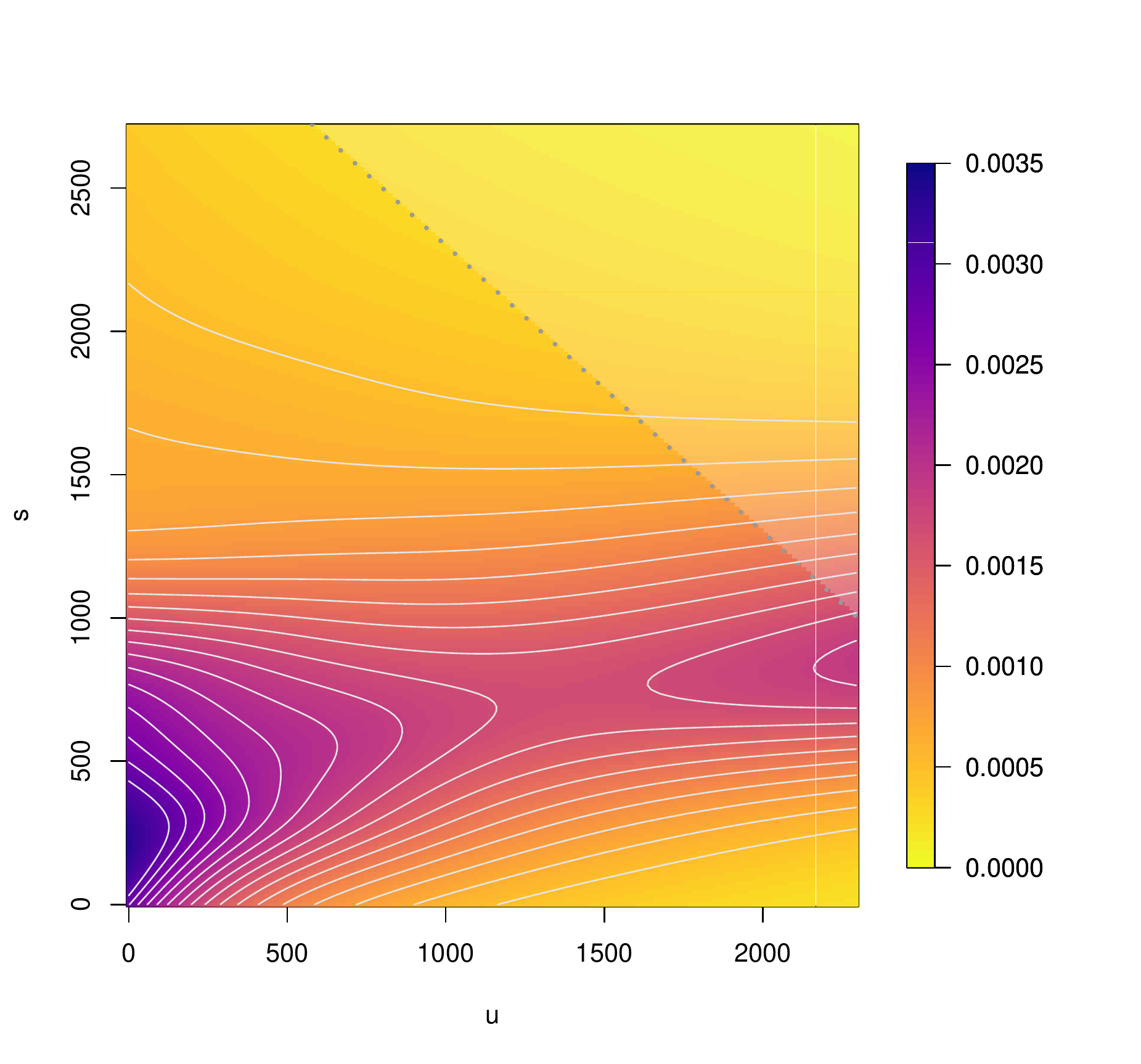}
\includegraphics[width=0.44\textwidth]{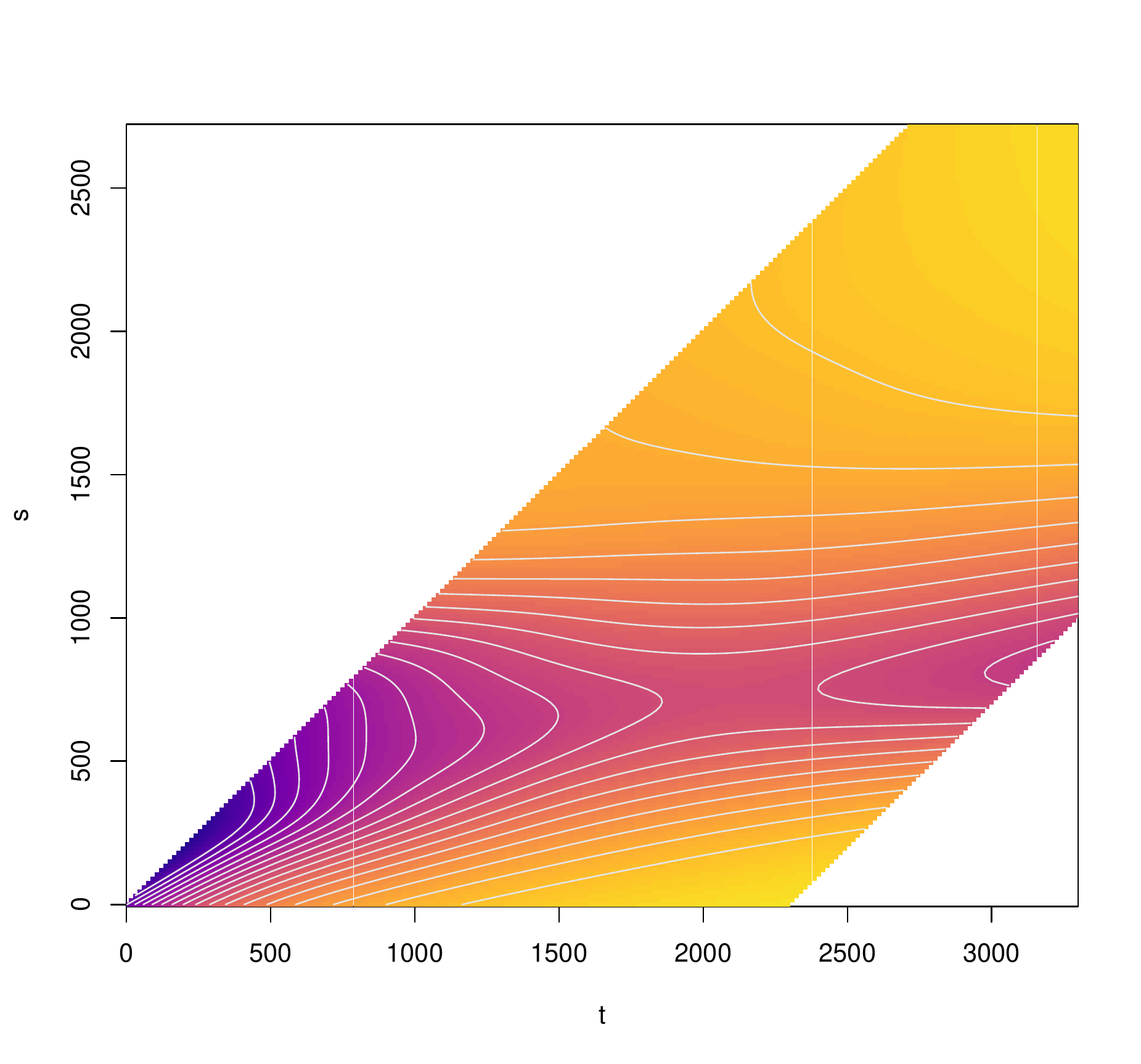}
\caption{Two-dimensional image of estimated hazard surface in $(u,s)$-plane (left) with extra\-polated area in top right corner indicated, and in $(t,s)$-plane (right).} \label{fig:colon2Destim}
\end{figure}

Figure~\ref{fig:colon3Destim} shows the resulting hazard surface $\hat{\breve \lambda} (u,s)$. A corresponding image plot, both in $(u, s)$ and $(t, s)$ coordinates is given in Figure~\ref{fig:colon2Destim}. In the $(u,s)$-plane in Figure~\ref{fig:colon2Destim} we marked the area in the top right where the surface is extrapolated beyond the data. In this application the extrapolation is unproblematic.

The images show that the hazard of death over $s$ changes with the timing of the recurrence. It reaches its highest level for early recurrences of the cancer, associated with early peaks in mortality. Peak mortality gradually decreases in level along $u$ while increasing its position on the $s$-axis. The pattern stabilizes for recurrence times $u$ at about 1200 days, which is about 3.25 years.

\begin{figure}[htbp]
\centering
\includegraphics[width=0.55\textwidth]{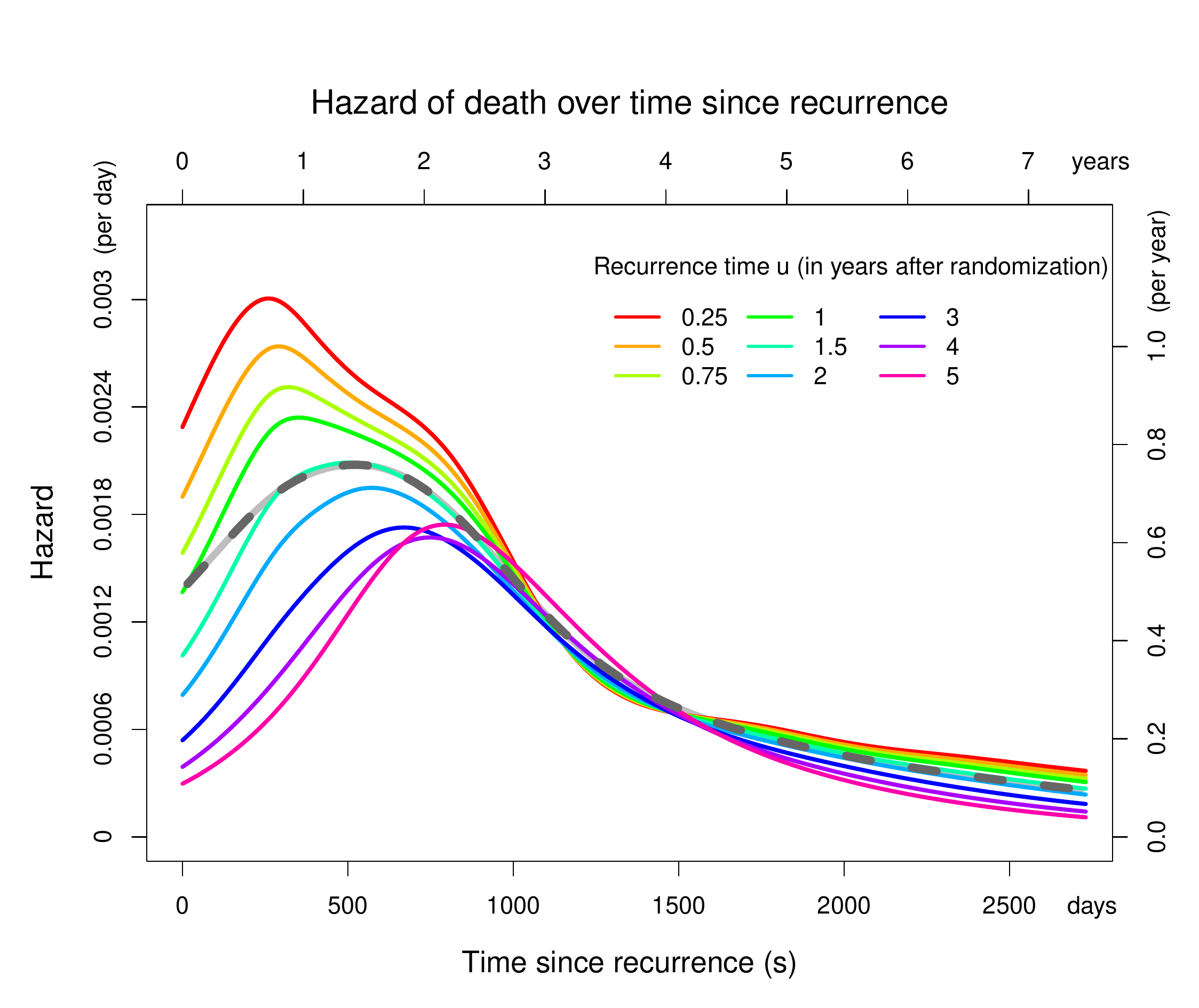}
\caption{Comparison of hazard of death over time scale $s$ for different times at recurrence~$u$. Dashed line marks the one-dimensional estimate obtained in Section~\ref{sec:estim1D}.}  
\label{fig:colon1Dcompare}
\end{figure}

To study this pattern further we cut the hazard surface at selected values of $u$ along~$s$. The resulting one-dimensional cutting lines are shown in Figure~\ref{fig:colon1Dcompare}. We also compare to the estimate that was obtained in Figure~\ref{fig:1Dhazard}, when the second time scale and consequently the interaction was ignored (wide dashed line). Considering only time since recurrence aggregates over the second dimension thereby missing the changing levels and variation in hazard shape. A simple way to capture variation in levels would be to 
add $u$ as a covariate in a proportional hazards specification, however, the two-dimensional hazard surface reveals the changing features of the hazard altogether.  

\begin{figure}[htbp]
\centering
\includegraphics[width=0.95\textwidth]{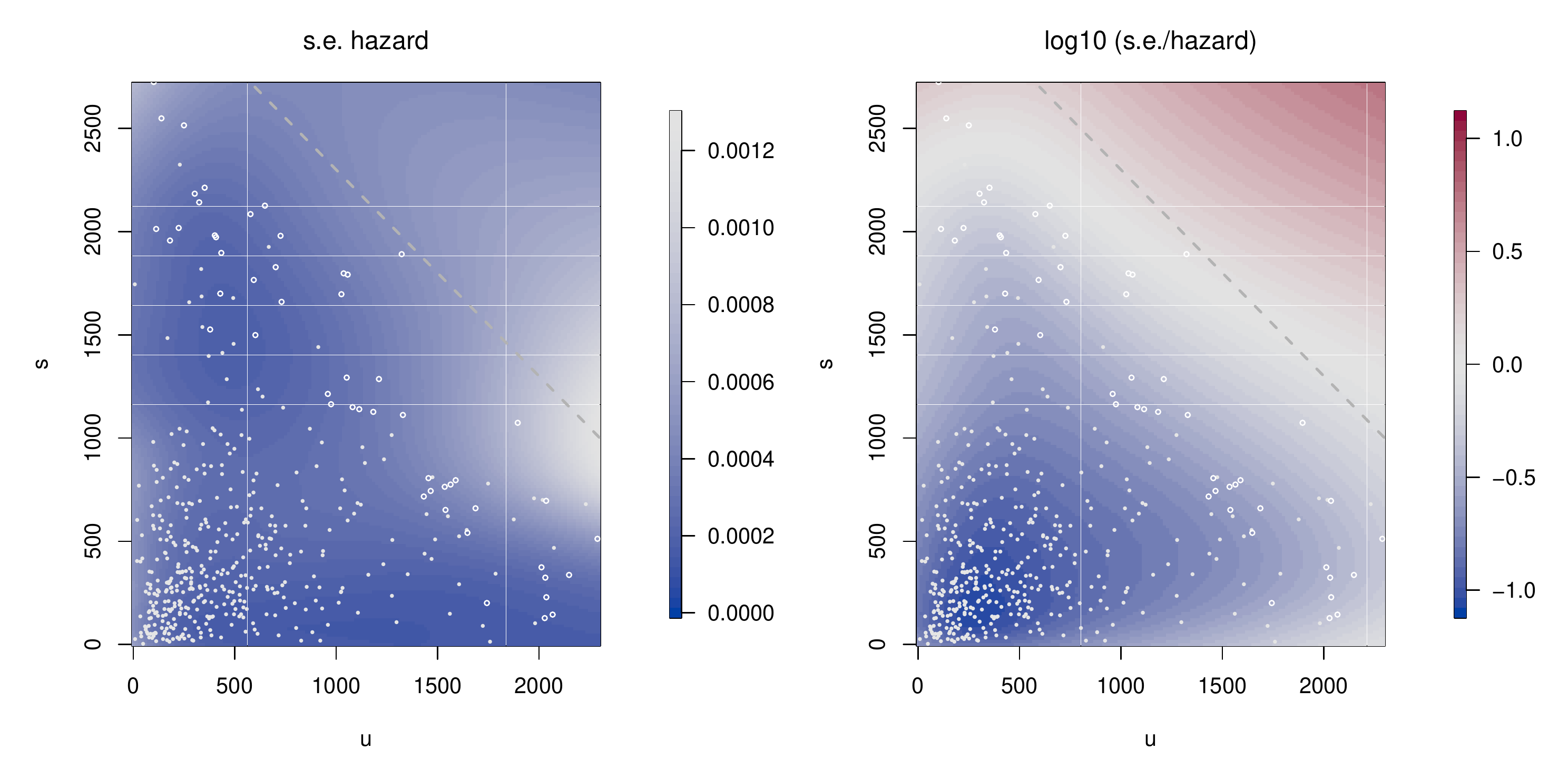}
\caption{Standard errors of $\hat{\breve\lambda} (u,s)$ (left) and standard errors relative to the hazard level on $\log_{10}$-scale (right). Original data added (see Figure~\ref{fig:colondescript}).} \label{fig:colon2Dse}
\end{figure}

The standard errors of the estimated surface are displayed in Figure~\ref{fig:colon2Dse}. The left panel shows the standard errors, while the right panel shows the standard errors relative to the hazard level on $\log_{10}$-scale. Naturally the uncertainty depends on the amount of information underlying the estimates. Therefore, to assess the trend in uncertainty, the underlying observations are added to the figures. Clearly, uncertainty is high in the extrapolation area and lowest where observations are densely packed, as one would expect.

\section{Proportional hazards regression with two time scales}
\label{sec:2DPH}

In the previous section we showed how to smooth a two-dimensional hazard surface without including additional covariates. The conventional proportional hazards (PH) model can also be specified in case of a baseline hazard that varies over two time scales:
\begin{equation}
\lambda(t,s; x ) = \lambda_0(t,s) \, \exp (x^\T \beta) = \breve \lambda_0(u,s) \,\exp(x^\T \beta),
\label{eq:hazardprop}
\end{equation}
where $\breve \lambda_0(u,s)$ is the baseline hazard surface as defined before and $x =(x_1, \ldots x_p)^\T$ is a $p$-vector of covariates and $\beta \in \mathbb R ^p$ are the corresponding regression parameters.
On the log-scale, if we model the baseline surface as in Section~\ref{sec:haz2D}, the overall predictor is linear:
\begin{equation}\label{eq:lpPH}
\eta(u,s; x) = \ln \breve \lambda  (u,s; x) = B(u,s)\,\alpha + x^\T \beta .
\end{equation}
What looks like a minor structural modification has considerable consequences for the computation though. 

The data contribution of a single individual $i$ are the entry and exit times on the two time scales, whether the exit was due to an event or censoring, and the vector $x_i =(x_{i1}, \ldots, x_{ip})^\T$ of covariates.
So in the tesselation, depending on the entry and exit times, each observation contributes positive at-risk times $r_{ijk}$ in the vertical bins in which the individual's $u_i$ is located (see Figure~\ref{fig:transformation}, right), and zero exposure elsewhere. Similarly, each individual contributes an event count $y_{ijk}$ of zero in all bins except the one where (s)he experienced an event. In the case of no covariates these $n$ sparsely filled event and exposure matrices, each of size $n_u\times n_s$, could be summed over $i$ and a single matrix for the total event counts $Y$ and exposure times $R$ represented the data, see equation~(\ref{eq:Poisson2}). Already in this case the regression matrix $B$ in (\ref{eq:tensor}) was of dimension $n_u n_s \times c_s c_u $, which led to the use the GLAM algorithm.

In the case of individual-specific hazards, induced by the covariates in vector $x_i$, this reduction is no longer possible and the `response' part of the data are three-dimensional arrays of size $n\times n_u\times n_s$.
So for the Poisson regression model for individual $i$ we have $Y_i = (y_{ijk})_{jk} $ and $R_i= (r_{ijk})_{jk}$
\begin{equation}\label{eq:Poisson3}
Y_{i} \sim \text{Poisson}(M_{i}), \qquad M_{i} = R_{i} \odot e^{E_{i}}, \qquad
E_{i} = (\eta_{ijk})_{jk} = B_u A B_s^\T + x_i^\T \beta,
\end{equation}
where the matrices $Y_i, R_i, E_i$ and $M_i$ above are of dimension $n_u\times n_s$, and $B_u A B_s^\T$ is the log-baseline-hazard surface, which is shared across all individuals, see (\ref{eq:2Dlp2}). 

Should we intend to write (and solve) the regression model in a flattened single matrix equation, then the design matrix would be of dimension $n n_u n_s \times (c_uc_s+p)$. It is obvious that careful matrix re-arrangements in GLAM style are needed to be able to handle the computations in reasonable time and with acceptable storage requirements. 
We defer the details of the matrix operations to Appendix~\ref{app:covars}.

Once the algorithm is set up, the penalized Poisson regression is again quick to converge, despite the size of the problem. Only the parameters $\alpha _{lm}$ for the smooth baseline surface will be penalized, while the regression parameters $\beta$ will remain unpenalized. The two smoothing parameters $\varrho _u$ and $\varrho_s$ are chosen by minimizing $\text{AIC}(\varrho_u, \varrho_s)$, as in Section~\ref{sec:haz2D}.

\section{Simulation Study}\label{sec:sim}
Before we apply the hazard regression model with two time scales to the colon cancer data in Section~\ref{sec:colonPH} we study the performance of the proposed approach in a simulation study. We consider several aspects that can affect the quality of the results: The complexity of the baseline surface, the sample size and the censoring and truncation pattern that influences the amount of information ultimately available in a sample. We start with exploring scenarios without covariates, PH regression is presented thereafter.

\subsection{Simulation settings}
\subsubsection{Hazard shapes}\label{sec:hazardshapes}
\begin{figure}[htbp]
\centering
\includegraphics[width=1.0\textwidth]{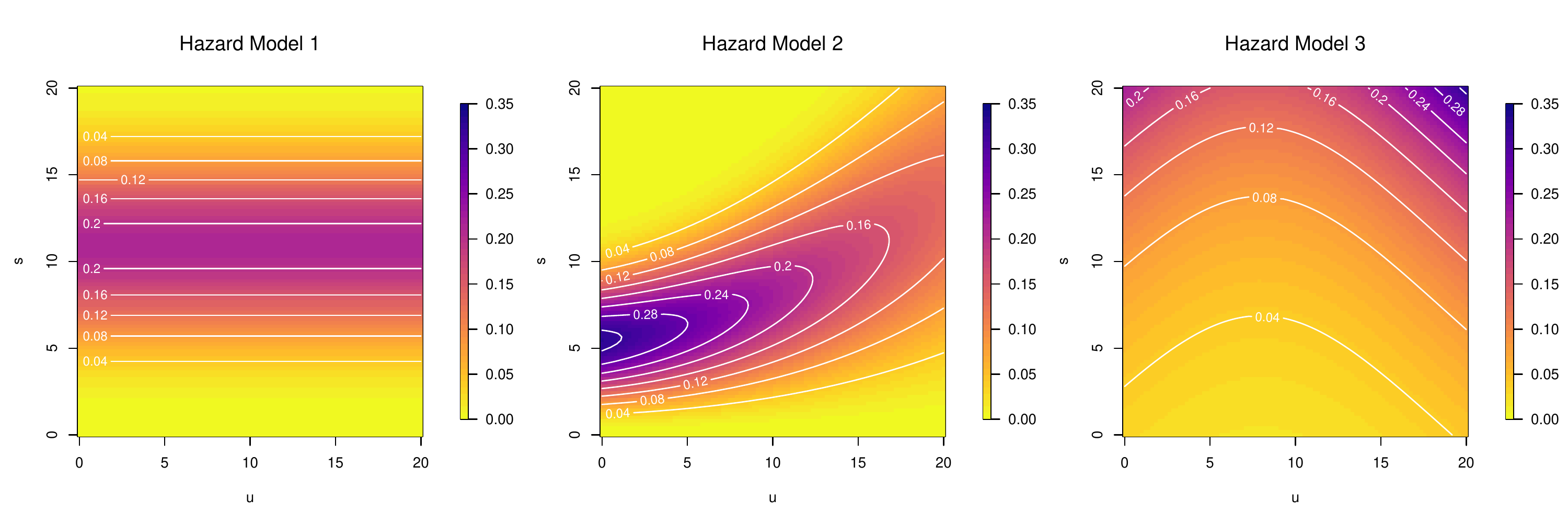}
\caption{Two-dimensional hazard shapes $\breve\lambda(u,s)$ used in the simulation study.}\label{fig:HM}
\end{figure}

For the two-dimensional hazard we consider three shapes of different complexity. They are presented as image plots $\breve\lambda(u,s)$ in Figure~\ref{fig:HM}. The first two specifications imply a unimodal hazard over $s$ that is changing (or not) with the value of $u$. In hazard model 1 (HM1) a single hazard shape persists for all values of $u$, while in model HM2 the location of the mode, the shape and the level of the hazard over $s$ smoothly changes with $u$. The third hazard HM3 is exponentially increasing along $s$ (Gompertz model) with parameters changing with $u$: $\breve \lambda (u,s) = a(u) \, e^{b(u) s}$. 
A unimodal hazard, as in HM1 and HM2, was found for the colon cancer data. An exponentially increasing hazard is regularly found in old-age disease incidence and mortality, so the last scenario HM3 intends to qualitatively capture such cases.
The detailed specifications of the hazard models are given in the Supplement. 

\subsubsection{Sample size and observation schemes}

For all three hazard models simulation of data sets proceeded in the following steps. For three different sample sizes, $n=300, n=500$ and $n=1000$, indiviudal values for $u$ were created randomly. In the simulation study the values $u_i$ were created from a uniform distribution on $(0, 20)$. 

Then, for each individual value $u_i$ a duration $s_i$ was simulated according to the hazard $\breve\lambda (u_i, s)$. The resulting triples $(u_i, s_i, t_i=s_i+u_i), i=1, \ldots, n$, form what we call the complete data (no censoring, no left-truncation).  On the complete data several observation schemes were imposed. 
In each scenario $S=100$ data sets were simulated.

Observation scheme A (OS\,A) imposed a maximum time $s_{\max}$ (set to 20) and observations with no event before $s_{\max}$ were right-censored. The observed events are hence found in $(0, 20) \times (0, 20)$ in the $(u, s)$-plane.

Observation scheme B (OS\,B) implements right-censoring along scale $t$: all individuals who have not experienced an event by $t=t_{\max}$ (set to 30) are right-censored at this value. As individuals differ in their values of $u_i$, the corresponding censored exit times will differ on the $s$-scale. For OS\,B events are found in the region $u\in (0,20)$ and $ s < t_{\max} - u$ in the $(u, s)$-plane. 
Both censoring mechanisms are independent of the process studied.

Observation scheme C (OS\,C) introduces some left-truncation. It operates on OS\,B and 20\% of the observations are randomly marked as late entries. Their entry times are drawn from a  uniform distribution on $(0, 6)$ and should they have experienced an event at time $s_i$ before their entry time, they are removed from the sample (left truncation). Hence datasets in OS\,C are generally smaller than the nominal sample size $n$.  As the hazard changes over $u$ in HM2 and HM3 the extent of left-truncation may vary across $u$.

Consequently, for each of the three hazard models we estimate $3 \times 3 = 9$ scenarios in the setting without covariates. 

\subsubsection{Regression models}
For the proportional hazards models we combine each of the above hazard surfaces and sample sizes with two covariates $x_1$ and $x_2$. Variable $x_1$ is quantitative and simulated from a standard Normal $N(0,1)$, $x_2$ is a centered binary  variable ($-0.5$ and $0.5$ with equal probability). The regression parameters are $\beta_1=0.5$ for $x_1$ and $\beta_2= 0.7$ for $x_2$. Once the individual values for the event times are created from the regression model, each complete data set is again submitted to the three observation schemes described in the previous section.

\subsection{Simulation results}

For reasons of space, we present here a synopsis of the simulation outcomes and defer a comprehensive documentation of the results to the supplementary material. 

In all settings the $(u, s)$-plane was split in bins of length $1$ along each axis. Cubic $B$-splines were used and the penalty order was $d=2$ along both rows and columns. The number of segments for the marginal bases was $12$ so that for the hazard $15^2 =225$ coefficients $\alpha _{lm}$ had to be estimated. The optimal values for the smoothing parameters $\rho_u$ and $\rho_s$ were determined by numerical minimization of $\mbox{AIC}(\rho_u, \rho_s)$. 

\begin{figure}[hbt]
\centering
Hazard Model 1

\smallskip
\includegraphics[width=0.97\textwidth]{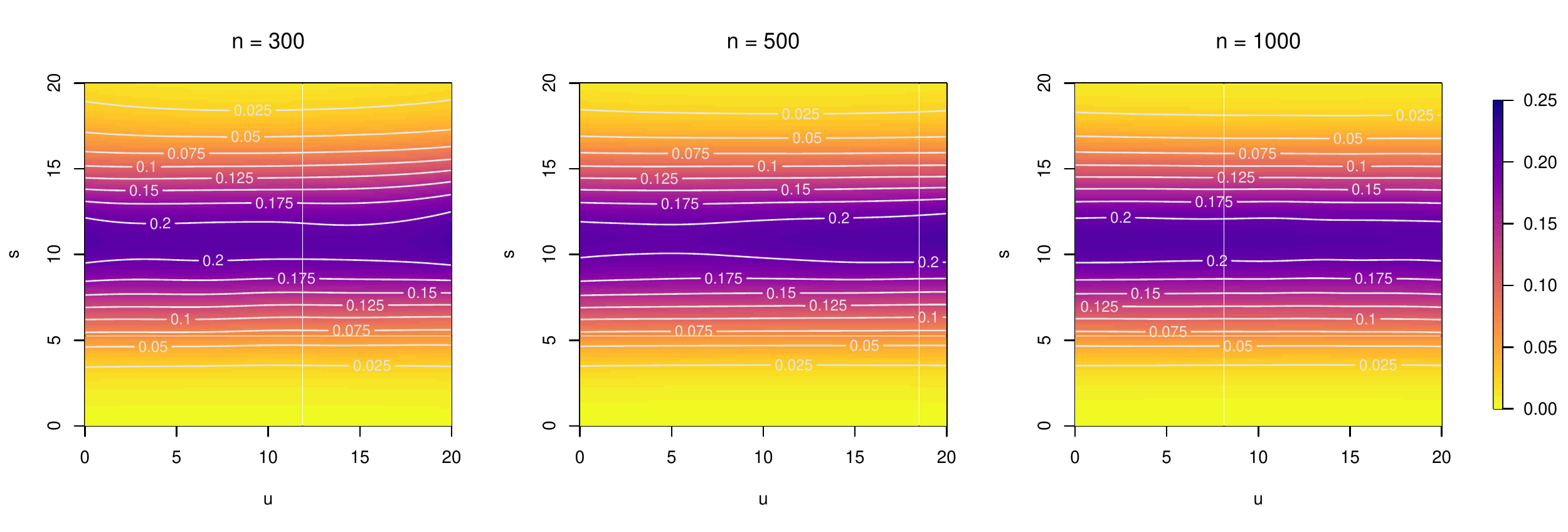}

\medskip
Hazard Model 2

\smallskip
\includegraphics[width=0.97\textwidth]{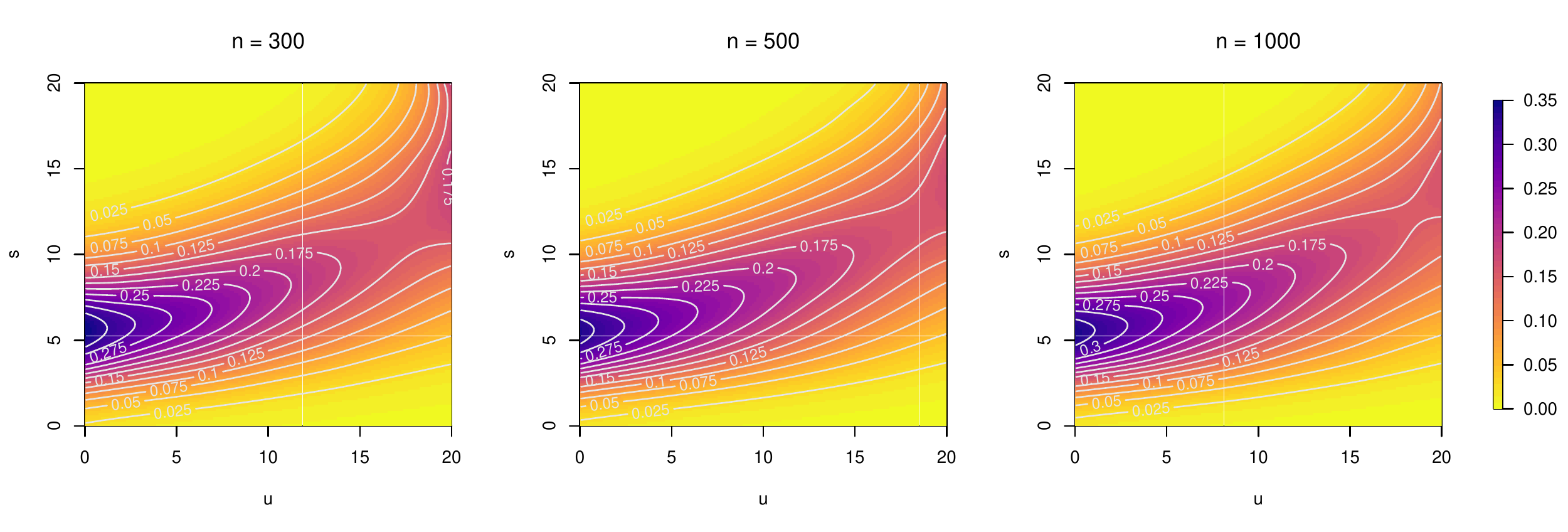}

\medskip
Hazard Model 3

\smallskip
\includegraphics[width=0.97\textwidth]{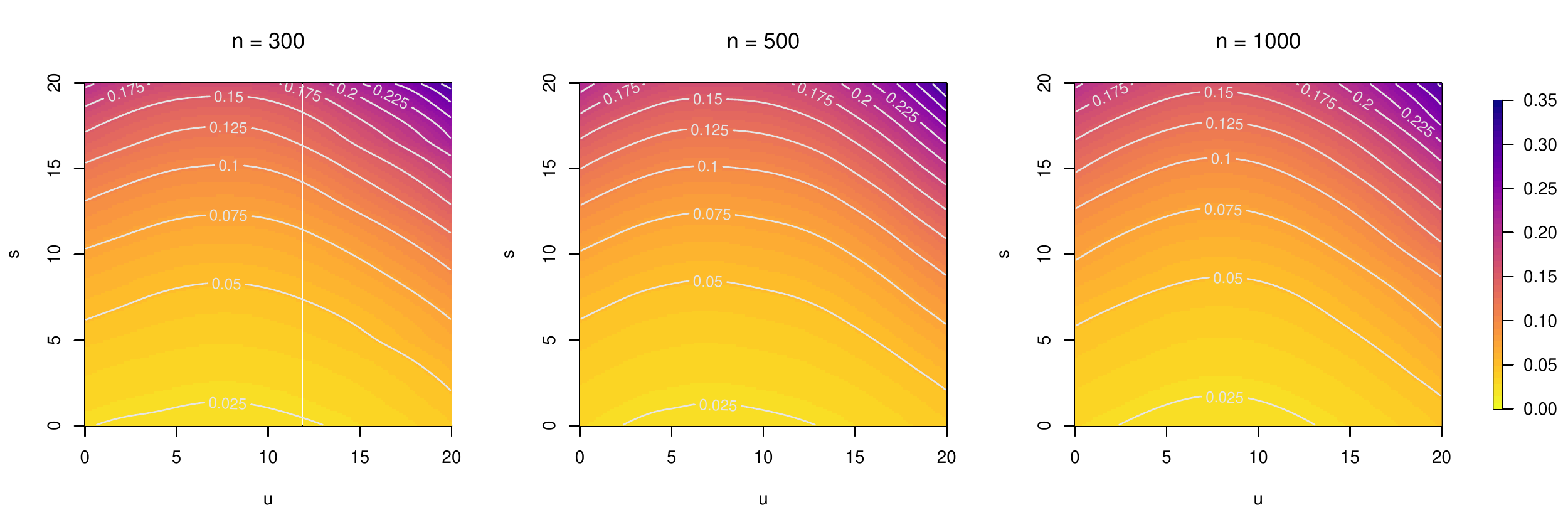}

\caption{Average of estimates for all three hazard models and sample sizes in observation scheme A (no covariates).}\label{fig:hazsummary}
\end{figure}

\begin{figure}
\centering
Hazard Model 1

\small Parameter $\beta_1$\hspace{16em}Parameter $\beta_2$
\includegraphics[width=\textwidth]{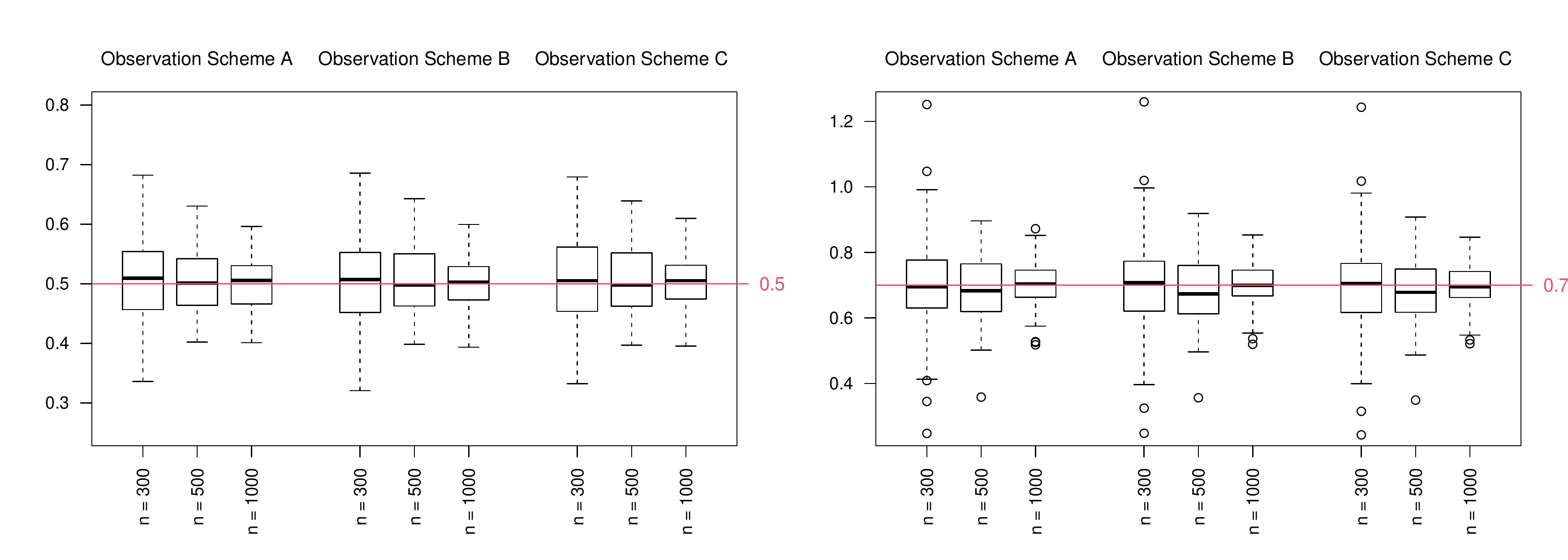}
\smallskip

\normalsize 
Hazard Model 2

\small Parameter $\beta_1$\hspace{16em}Parameter $\beta_2$
\includegraphics[width=\textwidth]{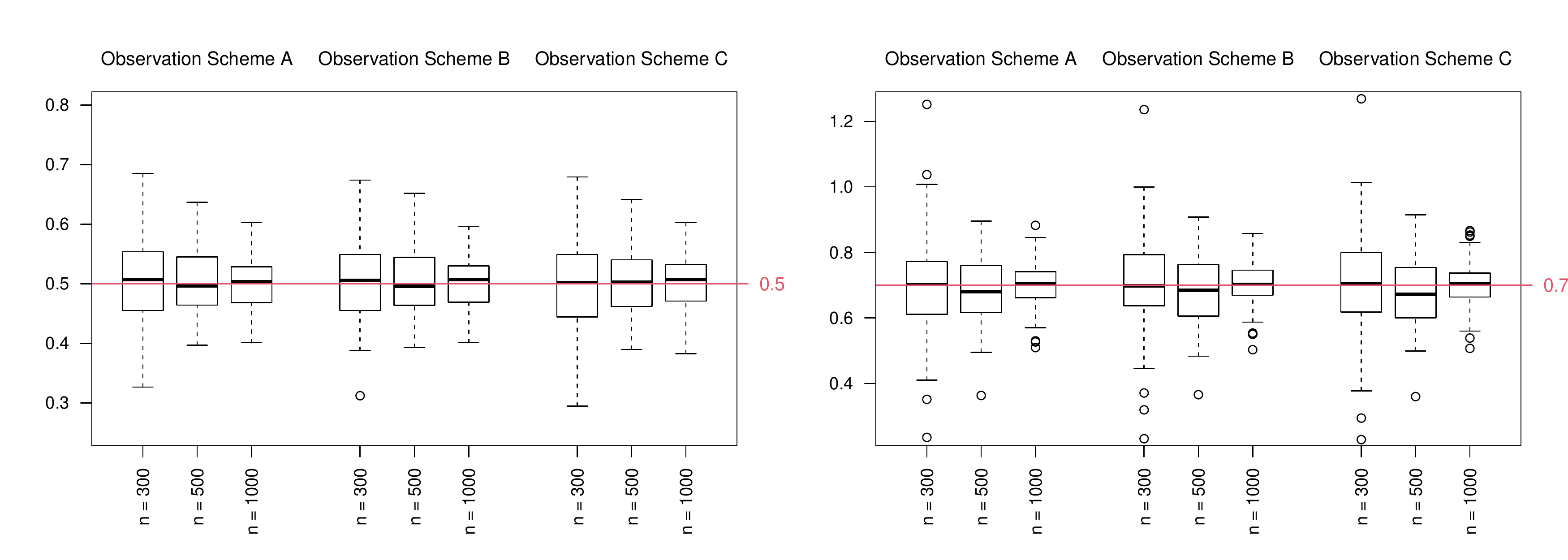}
\smallskip

\normalsize 
Hazard Model 3

\small Parameter $\beta_1$\hspace{16em}Parameter $\beta_2$
\includegraphics[width=\textwidth]{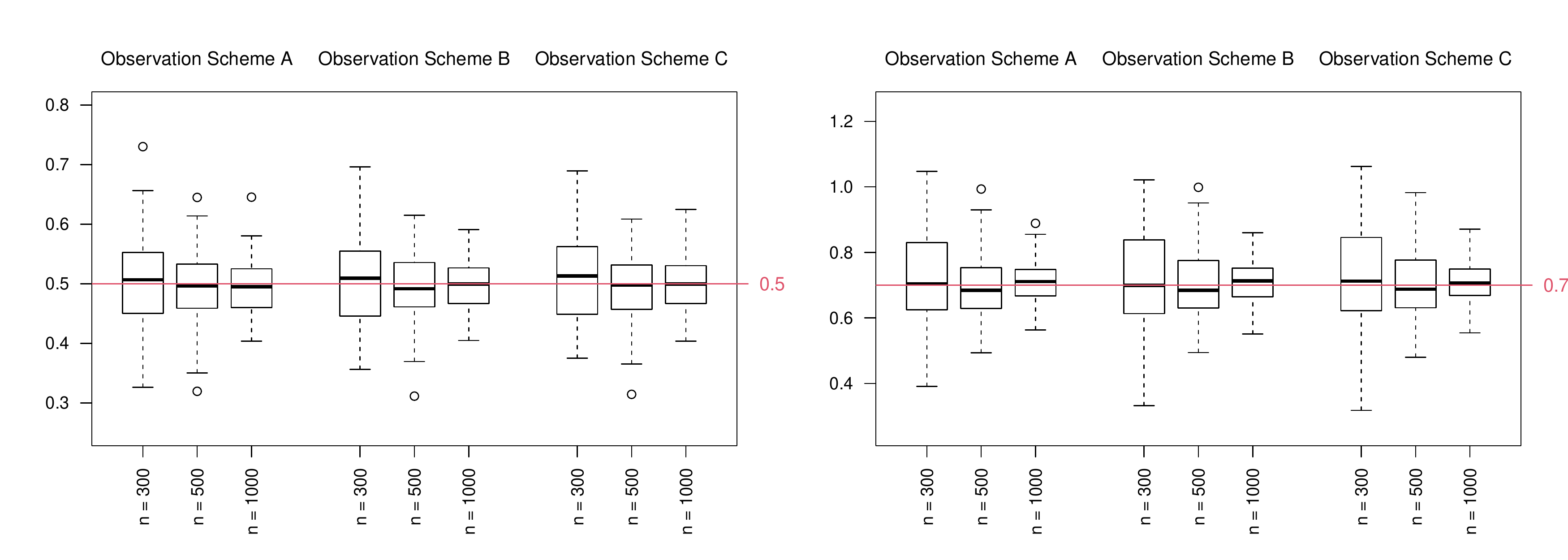}

\caption{Boxplots of the estimated regression parameters $\beta_1$ (left) and $\beta_2$ (right) for three baseline hazard specifications (top to bottom) in the simulation study. Within each sub-figure three observation schemes and three sample sizes are compared.}\label{fig:boxplotbeta}
\end{figure}

Figure~\ref{fig:hazsummary} shows the average estimated hazard $\hat{\breve\lambda} (u,s)$ across all three hazard shapes and for all sample sizes in observation scheme A (no covariates). Corresponding displays of the bias (mean difference between estimated and true hazard) as well as RMSE (root mean squared error) for this and other observation schemes are shown in the supplement.
Results for the regression parameters $\beta_1$ and 
$\beta_2$ are summarized, for all simulation settings, in Figure~\ref{fig:boxplotbeta}. 

As a general conclusion it can be said that the model captures the underlying structure well. The estimates are unbiased and variability decreases, as it should, with sample size. As for all multidimensional nonparametric smoothing methods there is some lower limit to the required sample size. The chosen value $n=300$ does not imply that this amount of data inevitably is required, since the amount of censoring and the way in which the observed events are scattered over the two dimensions also contributes to the estimation results. The more complex observation schemes do not affect the estimation results strongly. For a more detailed discussion and some practical recommendations see the supplement.
\vskip 5ex

\clearpage

\clearpage

\section{PH regression with two time scales for colon cancer data}\label{sec:colonPH}

We return to the colon cancer data introduced in Section~\ref{sec:colondata}. In Section~\ref{sec:colon2D} we estimated the hazard of death for patients with a relapse over two time scales but neglected additional covariates. Now we introduce information on the treatment (Levamisole, Levamisole+Fluororacil, reference = no treatment), sex of the patient (reference = female) and several binary indicators of disease severity (adherence to nearby organs, obstruction of colon by tumour, more than four positive lymph nodes) in a proportional hazards  regression model. The specification of the baseline hazard over the two time scales is identical to the one chosen in Section~\ref{sec:colon2D}. The resulting estimates are given in Table~\ref{tab:colonPH}, left.

\begin{figure}[htbp]
\centering
\includegraphics[width=\textwidth]{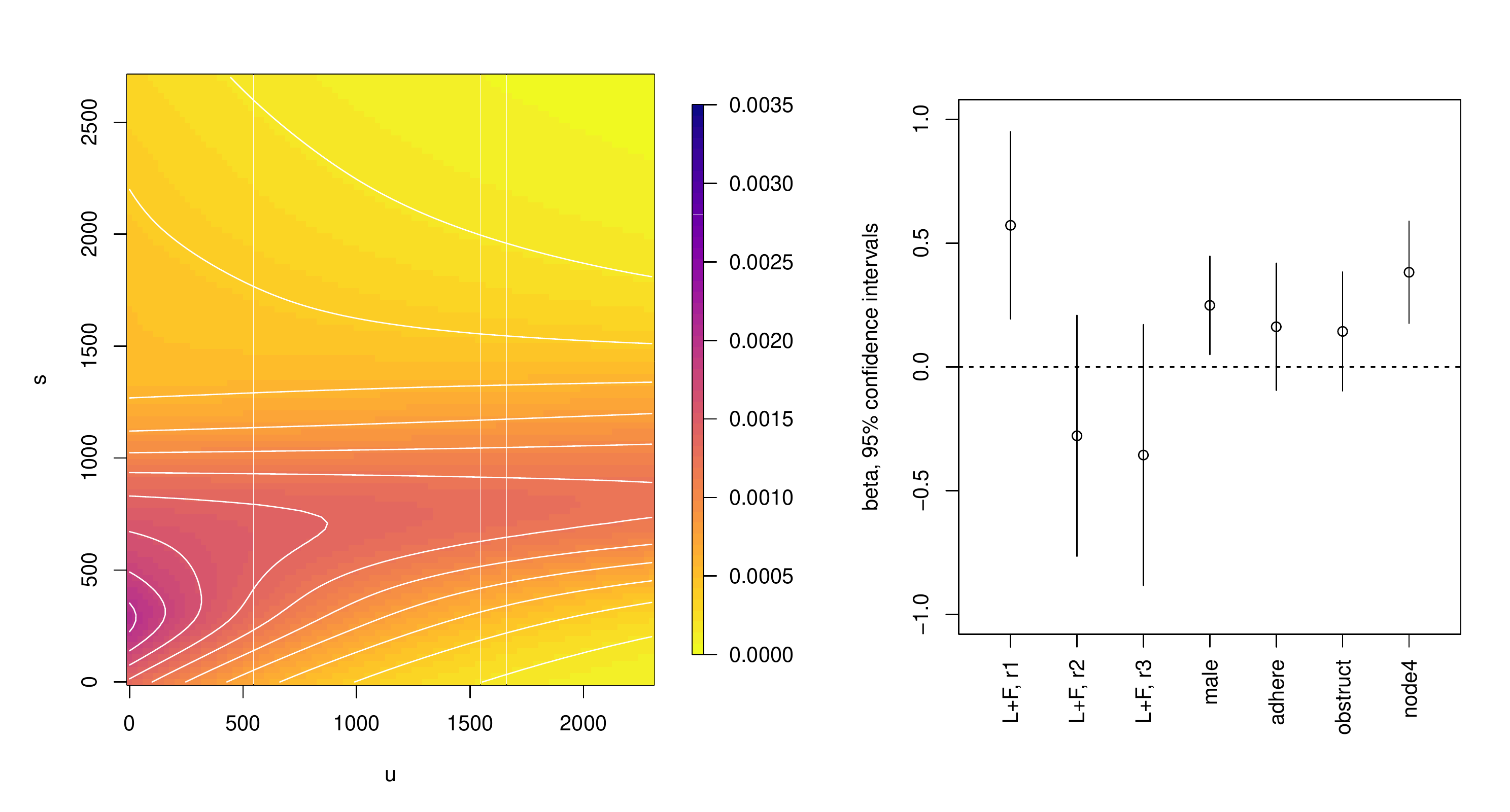}
\caption{Results of the PH model presented in Table~\ref{tab:colonPH}, right. Color coding as in Fig.~\ref{fig:colon2Destim}.}\label{fig:colonPH}
\end{figure}

\begin{table}[htbp]
\begin{center}
\medskip
\small
\begin{tabular}{lrr|lrr}
Covariate&$\hat\beta$ (s.e.)~~~~~& HR $e^{\hat\beta}$& Covariate&$\hat \beta$ (s.e.)~~~~~&  HR $e^{\hat\beta}$ \\
\hline
Lev&0.067 (0.115)&1.07&Lev+Fl, r1&0.572 (0.193)&1.77\\     
Lev+Fl&0.384 (0.130)&1.47&Lev+Fl, r2&-0.278 (0.248)&0.76\\
&&&Lev+Fl, r3&-0.356 (0.269)&0.70\\[2mm]
Male &0.254 (0.101)&1.29&&0.249 (0.101)&1.28\\
Adherence&0.154 (0.133)&1.17&&0.163 (0.131)&1.18\\
Obstruction&0.169 (0.122)&1.18&&0.144 (0.123)&1.15\\
Nodes $>$4&0.393 (0.105)&1.48&&0.383 (0.105)&1.47\\
\hline
ED baseline haz.$=9.8$&$\mbox{AIC}= 3073$~~~&&ED baseline haz.$=9.3$&$\mbox{AIC}=3073$~~~&\\
\end{tabular}
\end{center}
\caption{Estimated regression parameters, standard errors and hazard ratios (HR) for the colon cancer example. Left: PH model with treatments (vs.~control), sex and disease indicators. Right: Model with combined therapy (vs.~Lev and control) and effects that vary across the three thirds of time-to-recurrence distribution (r1, r2, r3). Other covariates as before. See also Figure~\ref{fig:colonPH}.}\label{tab:colonPH}
\end{table}
\citet{Moertel:1995} already noted that treatment by Levimasole alone did not show improvement over the control group. They also observed that the combined treatment, which was very successful in lowering the recurrence rate, was related to somewhat shorter survival times after relapse. To examine this result further, we estimate a second PH model in which the combined therapy (Lev+Fl), contrasted with the two other treatments (Lev and control), can have a different effect depending on the timing of recurrence, defined by the tertiles of the distribution of time to recurrence. The estimates are given in Table~\ref{tab:colonPH}, right and are also shown in Figure~\ref{fig:colonPH}. The increased risk of death for the combined therapy is only present for recurrences up to the first tertile, and the regression parameter is negative, though not significant, if recurrence occurred later. 
The color coding of the baseline hazard surface is the same as in Figure~\ref{fig:colon2Destim}. The baseline refers to low risk patients hence the lighter coloring. Including covariates reduces the complexity of the baseline. The effective dimension was $\mbox{ED}=11.2$ without covariates and is $\mbox{ED}=9.3$ for the PH model. Interaction still is present in the area of recurrence up to about two years, thereafter suggesting an additive model.

\section{Discussion and Outlook}\label{sec:discussion}

We demonstrated how two-dimensional $P$-spline smoothing can be deployed to model hazards that vary over two time scales. The binning of the data, which may be found uncommon at first glance, actually brings several advantages. It allows extreme flexibility for the hazard shape, the penalties provide smooth estimates, and the well-known iteratively weighted least squares iteration scheme is extended in a straightforward way to incoroprate the penalty. The good numerical properties of $P$-splines \citep[see][]{EilersMarx:2010} add to this. Second, in this way the model is in the class of generalized linear array models for which a suite of well-conceived algorithms is available that allow very efficient computations. All computations in this paper were performed using the companion R-package \texttt{TwoTimeScales} (see \textsl{https://github.com/AngelaCar/TwoTimeScales}), in which these GLAM algorithms are implemented.

In this paper the optimal values for the smoothing parameters were chosen by minimizing AIC. However, $P$-splines can be written as mixed models and optimal values of the smoothing parameters are then obtained from the estimated variances \citep[see][Appendix E]{EilersMarx:2021}. We plan to implement the mixed-model formulation also for the two time scales hazard model.

The analysis of the colon cancer data showed interaction between the two time scales, but often it will be of interest to explore whether a more simple model, such as an additive model for the log-hazard, fits the data sufficiently well. \citet{LeeDurban:2011} proposed ANOVA-type interaction models for spatio-temporal $P$-spline smoothing, which were extended further in \citet{LeeDurbanEilers:2013}, and we intend to adapt this idea to the two-dimensional hazard model.

In the data example the event times were known up to the day, so event times were exact. Right-censored and left-truncated information is included in a straightforward way in the approach. In practice, a common alternative observation scheme are interval-censored data, if patients are seen only at, more or less, regular intervals. In such an observation plan consequently neither the exact event times nor the exact at-risk times are known, however, they can be estimated employing an EM algorithm. This has been done for hazards with one time scale \citep{Gampe:2015} and we plan to extend this approach to the setting with two time scales as well.

Simple PH regression has been extended in many ways to overcome the relatively strict way how covariates affect the baseline hazard. Additive (rather than linear) predictors is one such extension, time-varying effects is another. In the current setting such extensions would fall within the scope of generalized linear additive smooth structures (GLASS), as coined by \citet{EilersMarx:2002}, with the extra complication of the additional two-dimensional baseline hazard. Smart arrangements in GLAM style certainly are needed for such extensions. This is a topic for future research.

\appendix
\section{Appendix}
\subsection{The GLAM algorithm}\label{app:GLAM}

The following description largely follows \citet{Currie:2006} and Appendix D in \citet{EilersMarx:2021}.

Section~\ref{sec:haz2D}  demonstrated that smoothing a two-dimensional hazard surface can be achieved by penalized Poisson regression. The IWLS algorithm (\ref{eq:IWLS}) requires to repeatedly solve the system
\begin{equation}\label{eq:IWLS2}
(B^\T\tilde W_\delta B + P) \,  \alpha = B^\T\tilde W_\delta B \, \tilde \alpha + B^\T ( y-\tilde \mu ),
\end{equation}
where the tilde indicates current values in the iteration.
Recall that the regression matrix $B= B_s \otimes B_u$ is the Kronecker product of the two marginal basis matrices and is of dimension $n_u n_s \times c_u c_s$, where $n_u$ and $n_s$ are the number of bins along the two axes.
Here we denote the $n_u n_s \times n_u n_s$ diagonal matrix of weights by $W_\delta$ to discriminate it from the matrix $W$ of dimension $n_u\times n_s$ that holds the diagonal elements of $W_\delta$ but arranged in the same manner as the event and exposure matrices $Y$ and $R$. 

The values of the log-hazard are arranged in the same way in matrix $\tilde E = [ \tilde \eta_{jk}]$ and $ \tilde M =[\tilde \mu_{jk}] = R \odot \exp ( \tilde E)$. 
In the Poisson model the weights are the $\mu_{jk}$, so $W=M$ and $W_\delta =\text{diag}(\text{vec}(M))$.

The inner products $B^\T\tilde W_\delta B$ and the right hand side of (\ref{eq:IWLS2}) have to be updated at each iteration. The penalty matrix $P$, which needs to be calculated only once, is given in~(\ref{eq:2Dpenalty}). 

If the number of $B$-splines along $u$ and $s$ is, say, $c_u=20$ and $c_s=20$ then $400$ coefficients need to be determined in (\ref{eq:IWLS2}). Solving systems of such sizes is not an obstacle anymore. The critical step is the formation of the inner-product matrix $B^\T\tilde W_\delta B $ with the tensor product matrix $B$. The size of $B$ is determined by the number of bins \textit{and} the number of coefficients. For example, if we use $n_u=n_s=100$ bins along each axis and $c_u=c_s=20$ $B$-splines, then $B$ has four million elements. 
To calculate the elements in $B^\T\tilde W_\delta B$ without explicitly forming the Kronecker product $B$ the following properties of Kronecker products are instrumental. 

First, we define the row-tensor $\phi(B)$ of a matrix $B$ with $c$ columns: 
\begin{equation}\label{eq:rowtens}
\phi(B) = (B\otimes \one _c ^\T) \odot (\one _c ^\T \otimes B),
\end{equation}
where $\one_c $ is a vector of ones of length $c$ (in our case $c=c_uc_s$). If $\tilde w$ is the vector of diagonal elements of $\tilde W_\delta$, it is straightforward to show that $B^\T\tilde W_\delta B $ and 
$\phi(B)^\T \tilde w $ contain the same elements only arranged in different ways: in $B^\T\tilde W_\delta B $ as a $c\times c$ matrix, in $\phi(B)^\T \tilde w $ as a vector of length $c^2$. Thus re-dimensioning of $\phi(B)^\T \tilde w $ renders $B^\T\tilde W_\delta B $. Note that  $\phi(B)^\T$ only has to be calculated once, while $B^\T \tilde W_\delta B $ needs to be updated whenever the weights in $\tilde w$ change.

Second, if $B$ is a Kronecker product, like $B= B_s\otimes B_u$, then one can also show that $\phi(B)^\T \tilde w$ and $\phi(B_u) ^\T \tilde W \phi(B_s)$, where $\tilde W$ is the $n_u \times n_s$ matrix of elements in $\tilde w$, likewise contain the same elements just arranged differently. The latter expression completely avoids forming the Kronecker product $B$.

Therefore proper re-arrangement of the elements of $\phi(B_u) ^\T \tilde W \phi(B_s)$ allows to recover all inner products in $B^\T\tilde W_\delta B$ without explicit calculation of $B$ and with considerably fewer multiplications. Re-arrangement operations are computatitionally cheap, so the procedure leads to substantial reductions in storage requirements and computation time.

Similarly, the elements of the right-hand side in (\ref{eq:IWLS2}) can be calculated without explicit\-ly forming the Kronecker product $B$ via 
\begin{equation}
B_u ^\T \left ( (Y-\tilde M) + \tilde M\odot\tilde E \right ) B_s   \notag
\end{equation} 
and re-dimensioning the $c_u \times c_s$ matrix as a vector of length $c_u c_s$.

To derive the variances of the linear predictor $\hat E = B_u \hat A B_s^\T $ the elements of the variance-covariance matrix $V =\text{Cov}(\alpha)$ of the coefficients, which is of dimension $c_u c_s \times c_u c_s$, are re-arranged in matrix $S$ of dimension $c_u^2\times c_s^2$ using array arithmetic. To obtain the $n_u n_s$ diagonal elements $\text{diag} ( B V B^\T)$ the multiplications with the tensor product $B$ again can be avoided. The same elements result from
\begin{equation}
\phi(B_u)\, S \, \phi(B_s)^\T , \notag
\end{equation}
fittingly arranged in the same way as the $n_u \times n_s$ matrix $\hat E$.

The GLAM procedure can be extended to more than two dimensions and the required re-dimensioning and rearrangement are provided in detail in \citet{Eilers:2006} and \citet{Currie:2006}.


\subsection{Computational details of the PH model in Section~\ref{sec:2DPH}}\label{app:covars}
The inclusion of individual-specific covariates prevents the aggregation of events and exposures across individuals and this enlarges the size of the problem.

For each individual we have two matrices, each of size $n_u \times n_s$, in which we collect the exposures and and event count in each of the bins. 
These matrices are extremely sparsely populated. Combining the matrices for all subjects, we
get two three-dimensional arrays, one for exposures, $R$ with elements
$r_{ijk}$, and one for events, $Y$ with elements $y_{ijk}$. Here $j$ and
$k$ index the two time scales and $i$ the subject. (We maintain the statistical notation $y_{ijk}$, in which the first index $i$ pertains to the individual. This is different from array indexing conventions in \texttt{R}.)

Furthermore, each individual contributes a $p$-vector of covariates $(x_{i1}, \ldots , x_{ip})^\T$ that are combined in the $n\times p$ matrix $X=[ x_{iv}]$.

The log-hazard for individual $i$ in bin~$j$ (of $u$-axis) and bin~$k$ (of $s$-axis) in the PH model is given as
\begin{equation}\label{eq:PHindiv}
\eta_{ijk} = \sum_{q=1}^{c_u}\sum_{r=1}^{c_s} b_{jq}{\check b}_{kr} \alpha_{qr} + \sum_{v=1}^p x_{iv} \beta_v .
\end{equation}

The $ b_{jq}$ is the value of the $q^{th}$ basis function along the $u$-axis, ${\check b}_{kr}$ is the $r^{th}$ element of the $B$-spline basis along $s$. 
Hence the first term in (\ref{eq:PHindiv}) expresses the tensor products for the baseline hazard which is shared across all individuals. The second term is the individual risk that raises/lowers the baseline for individual $i$.

Again, fitting the model boils down to penalized Poisson regression. And just as before, the size of the eventual penalized system of normal equations is not the problem but the size of the design matrix, and the weighted inner products based on the design matrix, pose the challenge.

In the design matrix of model (\ref{eq:PHindiv}) occur repeated tensor products of the $B$-spline bases (one for each subject) and repetitions of $X$ (one for each bin). If we denote this matrix by $C$ we can write it in the following way:

Let $B^+ = \one_n \otimes B$, where $\one _n$ is a column vector of $n$ ones and $B= B_s \otimes B_u$ is the tensor product matrix of the marginal bases. $B^+$ repeats the matrix $B$ $n$ times and therefore is of size $n n_u n_s \times c_u c_s$.
Similarily, define $X^+ = X \otimes \one_{n_u n_s} $ where $\one_{n_u n_s}$ is a column of ones of length $n_u n_s$. The matrix $X^+$ repeats each row of $X$ $n_u n_s$ times and therefore is of dimension $n n_u n_s \times p$.

With the definitions above the design matrix $C$ is $C= [B^+ \,|\, X ^+]$, which is of dimension \\ $n n_u n_s \times (c_u c_s + p)$. If we concatenate the parameter vectors $\theta = [\alpha ^\T \,|\, \beta ^\T ] ^\T $ we can write 
$$ \eta = C\theta $$
for the log-hazard, where $\eta$ is the appropriately vectorized $\eta_{ijk}$. Like before, the core challenge of the estimating algorithm is the computation of $G = C^\T V C$, where $V = \mathrm{diag}(\mu)$ and the vector $\mu$ results from $r \odot \exp(\eta)$ and $r = \text{vec}(R)$.

\noindent
Due to the particular structure of $B^+$ and $X^+$ the matrix $G$ can be partioned in
$$ G = \begin {pmatrix}
G_{11}~ & G_{12}\\
{G_{12} } ^\T& G_{22} 
\end{pmatrix}
$$
as follows:

\begin{itemize}
	\item As $B^+$ consists of $n$ stacked copies of $B$,  we have that
	 $G_{11} = n B^\T V B$, which is of dimension $c_u c_s \times c_u c_s$. Calculation of $G_{11}$ is performed using row tensors, see Section~\ref{app:GLAM}.
	\item As $ X^+$ contains $n_u n_s$ copies of $X$, we find that  
	$G_{22} = {X^\T} \, \mathrm{diag} (\underline v) \, X$, where $\underline v = (v_1, \ldots, v_n)^\T$ with 
	$ v_i = \sum_j\sum_k \mu_{ijk}$.	$G_{22}$ is of dimension $p  \times p$.
	\item $G_{12} = {B^+}^\T V {X^+}$. We re-dimension $V$ to the $n \times n_u n_s$ matrix $U$ and 
	then obtain $G_{12} = B^\T U^\T X$, which is of dimension $c_u c_s \times p$.
\end{itemize}   

\noindent
As only the parameters $\alpha$ in the baseline surface will be penalized, the penalty matrix (of dimension  
$c_u c_s \times c_u c_s$, see (\ref{eq:2Dpenalty}) ) is added to $G_{11}$ and inversion is done by using inversion formulas for partitioned matrices. 

\vskip 6ex
\bibliographystyle{chicago}
\bibliography{TTS1}

\end{document}